\documentclass[acmsmall,natbib=false,nonacm]{acmart}

\setcopyright{cc}
\setcctype{by}

\usepackage{amssymb}
\usepackage{multirow}
\usepackage{multicol}
\usepackage{svg}
\usepackage{booktabs}
\usepackage{subcaption}
\usepackage{hyperref}

\usepackage[style=acmnumeric,backend=biber,datamodel=acmdatamodel,giveninits=true]{biblatex}
\begin{document}

\title{Hardware-Level QoS Enforcement Features: Technologies, Use Cases, and Research Challenges}

\author{Oliver Larsson}
\orcid{0000-0003-0395-8313}
\affiliation{%
  \institution{Ume\aa{} University}
  \city{Ume\aa}
  \country{Sweden}
}
\email{olars@cs.umu.se}

\author{Thijs Metsch}
\orcid{0000-0003-3495-3646}
\affiliation{%
  \institution{Intel Advanced Technologies Group, Intel Corporation}
  \city{Munich}
  \country{Germany}
}
\email{thijs.metsch@intel.com}

\author{Cristian Klein}
\orcid{0000-0003-0106-3049}
\affiliation{%
  \institution{Ume\aa{} University}
  \city{Ume\aa}
  \country{Sweden}
}
\email{cklein@cs.umu.se}

\author{Erik Elmroth}
\orcid{0000-0002-2633-6798}
\affiliation{%
  \institution{Ume\aa{} University}
  \city{Ume\aa}
  \country{Sweden}
}
\email{elmroth@cs.umu.se}

\begin{abstract}
  Recent advancements in commodity server processors have enabled dynamic hardware-based quality-of-service (QoS) enforcement. These features have gathered increasing interest in research communities due to their versatility and wide range of applications. Thus, there exists a need to understand how scholars leverage hardware QoS enforcement in research, understand strengths and shortcomings, and identify gaps in current state-of-the-art research. This paper observes relevant publications, presents a novel taxonomy, discusses the approaches used, and identifies trends. Furthermore, an opportunity is recognized for QoS enforcement utilization in service-based cloud computing environments, and open challenges are presented.

\end{abstract}

\ccsdesc[500]{Computer systems organization~Cloud computing}
\ccsdesc[300]{Computer systems organization~Real-time systems}
\ccsdesc[500]{Software and its engineering~Memory management}
\ccsdesc[500]{Hardware~Semiconductor memory}

\keywords{Hardware QoS Enforcement, Cache Allocation, Resource Management, Cloud Computing, Congestion Control, Kubernetes}

\maketitle

{
    \small
    \copyright{} 2025 Copyright held by the owner/author(s). This is the author accepted manuscript of a paper published in \textit{ACM Computing Surveys (CSUR)}, open access under the Creative Commons Attribution (CC BY 4.0) license.  
    The final version is available at the ACM Digital Library: \url{https://doi.org/10.1145/3774317}.
}

\newcommand\revised[1]{#1}

\def\searchsize{3685}
\def\surveysize{78}

\section{Introduction}

Support for management and partitioning of shared system resources such as 
shared cache and memory bandwidth has recently become available in commercial 
off-the-shelf x86 server processors. The features, which we refer to as 
\emph{hardware-based quality-of-service} (QoS) \emph{enforcement features}, 
may be dynamically controlled at runtime  
using processor model-specific registers or an operating system-level interface. 
They enable systems
to enforce availability guarantees and mitigate the interference caused by 
shared resource congestion. The benefits of intelligently tuning such features have been shown 
in previous work to include increased system performance~\cite{lo_heracles_2015}, 
improved system utilization~\cite{patel_clite_2020}, 
and reduced power consumption~\cite{nejat_cooperative_2022}. 
However, to the best of our knowledge, there has been no previous attempt at
surveying the academic and industrial landscapes to understand the
applicability of these new features in different contexts. 

One such context of interest is that of general-purpose cloud
computing. The cloud-connected nature of day-to-day applications 
implies that performance issues of cloud-based web services now have a prevalent impact
on the daily lives of the general population in a way that was unthinkable a few decades ago. 
In fact, a 2023 US-based survey indicated that $48 \%$ of companies use
industry-tailored clouds with another $20 \%$ intending to make the transition to 
cloud-based solutions within 12 months~\cite{ballard_industry_2023}.
Latency-sensitive web services are often referred to as \emph{latency critical} (LC)
tasks, which is one of two broad categories cloud-based workloads may be classified into. 
The response times of LC services often directly affect the end-user experience and are therefore
monitored and kept to performance targets specified by service level objectives (SLOs).
\emph{Best-effort} (BE) jobs on the other hand are time non-critical, predictable workloads 
that may run whenever sufficient resources become available, commonly using a batch-format 
scheduling system. Thus, such jobs are 
well suited to improve server utilization by consuming resources left unused by LC services
based on current loads, a technique leveraged by several previous 
works~\cite{lo_heracles_2015, chen_parties_2019, nikas_dicer_2019, patel_clite_2020, papadakis_improving_2017, 
qin_locpart_2017,qiu_slo_2021,zhao_rhythm_2020,li_rldrm_2020,zhu_dirigent_2016, chen_avalon_2019}.
BE tasks such as machine learning training or big data 
analysis jobs are often data intensive and consequently disproportionately ``noisy'' 
users of shared system resources. Unless mitigated, this congestion may negatively affect the performance
of LC services~\cite{dasari_identifying_2013}.

The congestion of shared resources such as last-level cache (LLC) and memory bandwidth (MB) 
caused by the co-location of workloads on multicore systems 
has been shown in previous studies to reduce system
performance~\cite{dasari_identifying_2013, lofwenmark_understanding_2016, zhuravlev_addressing_2010}.
The importance of this becomes evident when considering the SLOs that web services often are required to meet.
Not reaching performance targets due to resource congestion may end up costly as BE jobs may need to be
canceled and re-queued. Alternatively, additional computing resources may need to be provisioned if permitted
by the underlying infrastructure platform. 
Hardware-based QoS enforcement features are designed to mitigate the issues of shared resource congestion.
However, hardware-based QoS enforcement is indirect in nature (see Section~\ref{sec:background}) 
and thus, intelligent tuning is required to achieve 
predictable and performant service operations in cloud environments. 

Although intelligent management of shared resources has been shown to improve workload performance
and energy efficiency, popular hypervisors and container orchestration engines still lack 
support for autonomous management of QoS enforcement on supported platforms. 
We thus identify an opportunity for further performance improvements in general-purpose 
cloud computing environments, but also question why the adoption of QoS enforcement
has not yet been made widespread despite the promises of improved performance.
Consequently, there exists a need to review the approaches proposed by researchers
that utilize hardware-based QoS enforcement features. We review published research 
and map how the features provide value and the problems they help solve. Furthermore, we propose
a novel taxonomy and investigate
the contexts in which such features are discussed such that 
gaps in current state-of-the-art research and areas of feature application may be identified. 
Additionally, a discussion of the reasons behind the lack of autonomous QoS enforcement
management solutions available in widespread workload management systems is presented.

\subsection{Our Contributions}

In this survey, we perform a thorough review of published research that integrates hardware-based 
QoS enforcement features as an integral part of their proposals. 
Thus, a complete scope of the currently identified applicability of features may be outlined. 
In summary, we make the following contributions:
\begin{enumerate}
    \item We survey a search space of $\searchsize$ articles and identify $\surveysize$ papers that
        fit the inclusion criteria defined for our review.
    \item We present a novel taxonomy of the included state-of-the-art research that details approaches
        and techniques that leverage hardware QoS enforcement.
    \item We discuss the ways hardware QoS enforcement features are utilized and identify reasons why 
        they have not reached widespread adoption in autonomous cloud computing environments.
    \item We detail gaps in current state-of-the-art research and present opportunities for future
        investigations.
\end{enumerate}

\revised{Our proposed taxonomy addresses a significant gap in the current literature. 
It serves as a foundation for understanding the state-of-the-art 
research and provides a novel overview of the applicability of hardware-based 
QoS enforcement features across various contexts. Additionally, we identify a 
lack of adoption of QoS enforcement in general-purpose cloud computing. 
By analyzing the literature, we aim to understand the common limitations and 
challenges that hinder the practicality and widespread adoption of QoS 
enforcement in the cloud. To this end, we present trends in the 
research landscape and identify challenges that need to be addressed to
facilitate wider adoption of QoS enforcement in autonomous cloud computing environments. }

\section{Hardware-based QoS Enforcement Features and their Origins} \label{sec:background}

The rapidly increasing number of cores available on a single multi-core processor has 
enabled larger numbers of workloads to be co-located on a single machine. While this has
its advantages, it puts increasing pressure on the resources shared among the 
cores of a system. 
Previous studies \cite{dasari_identifying_2013, zhuravlev_addressing_2010, lofwenmark_understanding_2016} 
have identified the main sources of performance interference in multi-core systems. 
\citeauthor*{dasari_identifying_2013}~\cite{dasari_identifying_2013} identified several main sources of
unpredictable delays due to resource interference. They conclude that shared main memory, shared cache, and
shared interconnection networks are the main sources of unpredictability. These claims are supported by 
\citeauthor*{lofwenmark_understanding_2016}~\cite{lofwenmark_understanding_2016}.
Furthermore, \citeauthor*{zhuravlev_addressing_2010}~\cite{zhuravlev_addressing_2010} found in their 
2010 study that shared cache, main memory bus, and memory access controller 
congestion were all prominent sources of interference.  

To address the problem of shared cache interference, \citeauthor*{iyer_cqos_2004} introduced the
notion of QoS in the context of shared cache management in 2004~\cite{iyer_cqos_2004}.
They theorized an early cache partitioning approach that was evaluated using cache simulators. 
While this proposal was rudimentary, it was developed 
upon and refined proposals of cache and memory partitioning approaches were 
later published~\cite{iyer_qos_2007}. Simultaneously, low-overhead cache monitoring 
solutions were also theorized~\cite{zhao_cachescouts_2007}. 

In more recent years, chip manufacturers have started to implement cache and memory bandwidth 
management features in hardware. 
Intel refers to these
features as Intel Resource Director Technology (RDT)~\cite{intel_corporation_intel_2019}, 
while AMD calls their implementation AMD Platform Quality of Service Extensions 
(QoSE)~\cite{advanced_micro_devices_inc_amd64_2022}. The features allow practitioners to 
tune hardware to the specific needs of their workloads and thus achieve improved efficiency 
through reduced interference.

\revised{This article's scope focuses on x86-based implementations of hardware-level QoS enforcement. 
However, outside the domain of x86 processors, similar features have been implemented in 
the ARM ecosystem by Cavium (now Marvell)~\cite{wang_swap_2017} as proprietary features of their processors. 
More recently, official ARM support for QoS enforcement features has been made available 
through the Memory System Resource Partitioning and Monitoring (MPAM) 
technology~\cite{arm_limited_arm_2024}. Additionally, efforts have been 
made to introduce similar features to RISC-V platforms~\cite{zhang_labeled_2022}.}

The remainder of this section discusses the causes of performance unpredictability originating in shared resource
congestion. This requires an insight into the architecture of modern multi-core processors, 
as well as how they implement the specific resource of interest. Additionally, the inner 
workings of the dynamic resource control mechanisms of modern x86 processors are discussed.

\subsection{Classes of Service}

The control mechanisms of hardware-based QoS enforcement features use a concept 
called classes of service (COS or CLOS). This concept is grounded in hardware design
and dictates how resource control rules are applied to the workloads of a system. 
Every logical core is assigned to one, and only one, COS at any one time. Resource control
rules are set at the COS level, which will be enforced upon any core in the COS.
As the system starts up, all cores are assigned to \texttt{COS0}, the default COS~\cite{yu_24_nodate}.
They may later be assigned to different COSs dynamically at runtime. The total number of available COSs is
dependent on the processor model and thus changes between systems.

\subsection{Shared Cache Congestion} \label{sec:cache_congestion}

Traditional von Neumann architecture systems have processing and memory units 
physically separated~\cite{arikpo2007neumann}. Thus, the performance of a system becomes highly 
dependent on the performance of the interconnect between the two physical 
units, which has become known as the \emph{von Neumann bottleneck}~\cite{arikpo2007neumann}. 
Over the past few decades, advances in processing technology have increasingly outgrown that of
memory technology~\cite{hennessy_computer_2011} and
this difference in performance growth is referred to as the 
\emph{memory wall}~\revised{\cite{wulf_hitting_1995}}, which has prompted the invention of many 
mitigating techniques.

One such technique is to use a \emph{cache}. A cache is a small, high-performance 
memory package that is typically embedded in the CPU die. As the cache is
substantially faster than the main memory, the processor may place data currently
worked on in the cache and thus spend less time waiting for the data it needs from the main 
memory~\cite{brett_memory_2016}. 

\begin{figure}[htbp]
    \centering
    \small
    \includegraphics{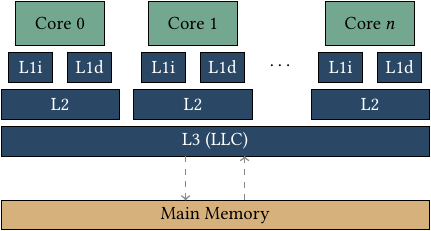}
    \caption{Illustration of a typical three-\revised{level} cache architecture.}
    \label{fig:cache}
    \Description{Illustration of a typical three-level cache architecture.}
\end{figure}

The caches of modern multi-core processors are so-called multi-\revised{level} caches. Each \revised{level}
of cache is larger, but slower than the last. Typically, modern caches consist of three
\revised{levels}. \revised{Level} 1 (L1) is the smallest and fastest \revised{level}. Each physical core 
has a separate L1 cache split into two parts, a data cache (L1d), and an instruction 
cache (L1i). The next \revised{level}, \revised{level} 2 (L2), is similarly constructed so that each
physical core has its own instance. However, there is no distinction between data and
instruction memory in the L2 cache. \revised{Level} 3 (L3) is shared among multiple cores, 
often all cores on a chip. This means that there may be inter-core 
interference as the workloads of different cores compete for cache space. 
The highest numbered level of cache on a system is commonly known as the 
last-level cache (LLC). On a three-level cache system, LLC refers to the L3 cache. 
An overview of a three-\revised{level} cache architecture is illustrated in \figurename~\ref{fig:cache}.

Due to the cache's small size, it is inevitable 
that memory will have to be read from the main memory eventually as working set sizes outgrow
the available cache. AMD's EPYC Genoa server processors with 3D V-Cache technology provide 
up to $1152$ megabytes of LLC cache while supporting up to 6 terabytes of main 
memory per socket~\cite{advanced_micro_devices_inc_amd_2023} and thus, LLC accounts for only $0.02 \%$ of 
the main memory size.
When memory required by the processor is not already present in the cache, a \emph{cache miss} has occurred
and a delay penalty is incurred while waiting for the required data~\cite{dasari_identifying_2013}. 
In a 2022 study by \citeauthor*{velten_memory_2022}, it was found
that L3 access incurred a penalty of 39 cycles while waiting for main memory access 
took 221 cycles~\cite{velten_memory_2022}\footnote{
    L3 and main memory access times are highly dependent on hardware~\cite{velten_memory_2022}. 
    The numbers presented were recorded on a system equipped with two AMD EPYC 7702 (Rome) processors.
}. Thus, accessing data
not present in any level of cache results in a $5.67 \times$ increase in access time compared to accessing
data present in the shared L3 cache.

Modern data centers spend an increasing proportion of computational resources 
on data-intensive workloads such as machine learning (ML) training, ML inference, 
video processing, and audio processing. The size of the data sets used by these processes 
means that they are particularly destructive to cache integrity, meaning more time 
is spent waiting for data from the main memory. Given the shared nature of the LLC, 
this affects all workloads of a system and is known as \emph{shared cache congestion}.
Such congestion may take the form of \emph{inter-task} interference when a process 
replaces cache lines of another process on the same core, or \emph{inter-core} interference when
cache lines of a process running on another core are overridden~\cite{lofwenmark_understanding_2016}.

\subsubsection{Shared Cache Partitioning} \label{sec:cache_allocation}

Hardware support for cache partitioning has been implemented by chip manufacturers  
in recent years. Tuning the cache partitioning scheme may reduce
the impact of shared cache congestion on workloads. As we will see 
in this survey, it has been repeatedly shown that cache partitioning 
may be used to great effect when optimizing the workloads of a system. 

\begin{figure}[htbp]
    \centering
    \small
    \includegraphics{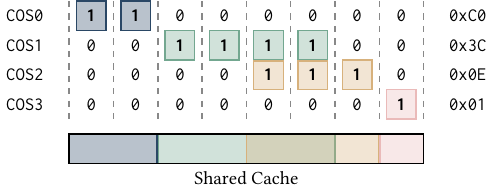}
    \caption{Illustration of a cache partitioning scheme for a CPU supporting $8$ bits of 
        control ($0_{16}$ to $\textrm{FF}_{16}$). Here, \texttt{COS0} has exclusive access
        to $2/8 = 1/4$ of available cache, while \texttt{COS1} and \texttt{COS2} share parts of
        their allocation.}
    \label{fig:cache_allocation}
    \Description{Illustration of a cache partitioning scheme for a CPU supporting $8$ bits of 
        control ($0_{16}$ to $\textrm{FF}_{16}$). Here, \texttt{COS0} has exclusive access
        to $2/8 = 1/4$ of available cache, while \texttt{COS1} and \texttt{COS2} share parts of
        their allocation.}
\end{figure}

Configuration of the cache allocation scheme for a COS is done by setting a bitmask,
the size of which is dependent on the specific CPU model. 
Typically, each bit
corresponds to a separate hardware \emph{cache way}~\cite{sohal_closer_2022}, but this is not necessarily the case~\cite{yu_24_nodate}.
A CPU supporting $n$ bits of 
control may have a COS assigned $1/n$ or 
more\footnote{COSs assigned more than one such share must on Intel platforms have 
all ``1'' bits in a contiguous block due to hardware limitations~\cite{intel_corporation_intel_2023}.}
of the available cache. Allocation bitmasks may overlap between COSs. If so,
parts of the cache are shared between two or more classes. 
An illustration of a
cache partitioning configuration may be observed in \figurename~\ref{fig:cache_allocation}.

When moving a core to a new COS, the new rules only apply 
to future cache allocations. Cache lines already present in the old partition 
may continue to be used until evicted~\cite{yu_24_nodate}. 

Some hardware additionally supports the \emph{code and data prioritization} (CDP) extension
of the cache allocation interface. When enabled, CDP permits the configuration of 
separate cache allocation bitmasks for instructions and data.

\subsection{Shared Memory Congestion}

\begin{figure}[htbp]
    \centering
    \small
    \includegraphics{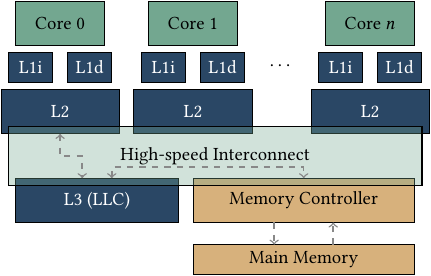}
    \caption{Illustration of the shared memory bus of a modern 
    Intel Xeon Scalable processor~\cite{mulnix_intel_2022}.}
    \label{fig:memory_bus}
    \Description{Illustration of the shared memory bus of a modern 
    Intel Xeon Scalable processor.}
\end{figure}

Although techniques such as caching mitigate the von Neumann bottleneck described in 
Section~\ref{sec:cache_congestion}, they do not remove the need to transfer data to and
from the main memory. The interconnect between the CPU and the main memory, the \emph{memory bus}, 
is limited in bandwidth and shared by all tasks on a non-uniform memory access (NUMA) node. 
Therefore, memory requests
may be required to do additional waiting until sufficient capacity to handle the
request is available~\cite{pellizzoni_worst_2010}. This may result in workloads 
of low to medium memory intensity 
becoming drowned out given co-located workloads of high enough memory intensity. 
An abstract model of the interconnect between the cores, LLC, and main memory
is illustrated in \figurename~\ref{fig:memory_bus}.

Memory controllers commonly implement a variation of a first-come-first-served (FCFS) request
scheduling algorithm that prioritizes row hit requests~\cite{rixner_memory_2000}. This 
prioritized queue of requests in the memory controller influences the memory access times of 
processes~\cite{kim_bounding_2016}.

\subsubsection{Shared Memory Bandwidth Limiting}

Partitioning of the memory bus is available on supported 
hardware, which functionality is provided by a
programmable request rate controller implemented in chip hardware.
The feature is referred to as \emph{memory bandwidth allocation} (MBA) by
Intel~\cite{intel_corporation_intel_2023} and the Linux kernel 
documentation~\cite{yu_24_nodate}. However, a more descriptive name is \emph{bandwidth 
limitation} as used by AMD~\cite{advanced_micro_devices_inc_amd64_2022}. 

Bandwidth limitation provides the ability to dynamically throttle and delay
memory requests from specific processor cores. Typically, 
a memory-intensive process would have its memory requests limited so that
other workloads become less affected by the high bandwidth usage.
Bandwidth limits may be specified as a percentage (Intel only), 
or as a maximum bandwidth in megabytes~\cite{yu_24_nodate}. 

\begin{figure}[htbp]
    \centering
    \small
    \includegraphics{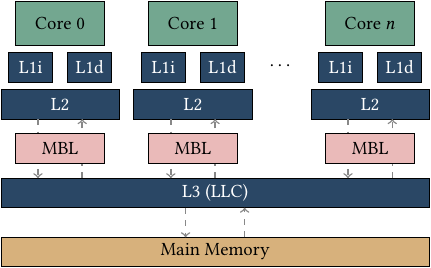}
    \caption{Illustration of memory bandwidth limiting (MBL) impact on the memory request chain.
        Bandwidth limiting affects all requests that miss the L2 cache, which includes potential L3 hits.}
    \label{fig:memory_limiting}
    \Description{Illustration of memory bandwidth limiting (MBL) impact on the memory request chain.
        Bandwidth limiting affects all requests that miss the L2 cache, which includes potential L3 hits.}
\end{figure}

Throttling is applied to all requests that miss L2 cache, the 
last \revised{level} of cache specific to a single physical core. This includes 
requests that may hit the shared L3 cache~\cite{xiang_emba_2019} as illustrated
by \figurename~\ref{fig:memory_limiting}. Thus,
bandwidth limitation is the most effective at throttling workloads
that are inefficient at utilizing the L3 cache. Throttling is applied 
at the level of physical cores and any core utilizing simultaneous multi-threading
will therefore use the maximum throttling value of its 
logical cores~\cite{herdrich_introduction_2019}.

\subsection{\revised{Indirectness of Hardware-Level QoS Enforcement}}

\revised{Many of the challenges of adopting hardware-level QoS enforcement features stem from 
the indirect nature of their design. Both cache partitioning and memory bandwidth
limiting (MBL) exhibit this indirectness. Let us first consider MBL. Any given 
workload will have certain memory bandwidth requirements that among other things
is dependent on other system workloads through congestion in shared cache. 
It is not possible to give direct priority to a workload to satisfy these requirements using MBL. 
Rather, other system workloads must be throttled to indirectly free up bandwidth for 
the workload in focus.}

\revised{In the case of cache partitioning, the problem of indirectness is less obvious.
Intuition may tell us that assigning a larger cache allocation to a workload will 
improve its performance. However, as was shown by 
\citeauthor*{chung_enforcing_2019}~\cite{chung_enforcing_2019}, more LLC is not always 
better. They found that the performance of a workload may be negatively impacted 
by an increased cache allocation. This is due to the fact that other co-located 
workloads now share a smaller portion of cache and may subsequently experience more
cache misses. This increase in cache misses result in increased congestion over
the memory bus, saturating its bandwidth. Thus, when the workload in focus
eventually requires data from the main memory, it may be delayed due to the 
congestion caused by other workloads which results in a net negative performance.
It is thus not possible to make assumptions about workload cache allocation 
without considering the indirect behavior of other co-located workloads.}

\subsection{OS Support for Hardware QoS Enforcement}

Management of QoS enforcement schemes is preferably managed through an operating system-level
interface. In Linux, this interface is known as the \emph{User Interface for Resource 
Control} (\texttt{resctrl})~\cite{yu_24_nodate}. \texttt{resctrl} enables the definition of 
resource allocation groups that dictate resource behavior. Logical processor cores or even
individual processes may be assigned to a resource group and will subsequently follow the 
allocation configuration of that group.
Resource allocation groups are mapped directly to hardware classes of service (CLOS or COS). 
Therefore, the number of available resource groups is limited and dependent on the CPU model. 
When individual processes are assigned to resource groups, the rules of the core executing 
that process are modified at the time of context switching such that the rules only apply to 
the correct process.

The resources available to any one process are determined by following a set of rules:
\begin{enumerate}
    \item If the process is assigned to a resource group, 
        the configuration of that group is used.
    \item If the process is in the default resource group but is 
        running on a processor core assigned to a resource group, 
        then the configuration of the core's group is used.
    \item The configuration of the default resource group is used.
\end{enumerate}

\subsection{A Note on Shared Resource Monitoring Features}

In addition to the enforcement features discussed, the Intel RDT and AMD QoSE feature sets
include monitoring support for shared cache and memory bandwidth. These monitoring features are,
similarly to their enforcement counterparts, accessible through the \texttt{resctrl} 
OS interface~\cite{yu_24_nodate}.

Due to the indirect nature of the mechanisms provided by QoS enforcement features, 
tuning behavior is highly dependent on the dynamics of the workloads 
of a system. Thus, actual behavior may be difficult to predict \cite{sohal_closer_2022, 
farina_assessing_2022, xiang_emba_2019} and it is therefore recommended that 
actual resource usage is monitored and taken into account by controllers~\cite{
intel_corporation_intel_2023}. Monitoring features therefore play a crucial role in effective
feature utilization as they allow controllers to observe changes in metrics resulting from
a tuning decision.

\section{Review Motivation} \label{sec:motivation}

By observation of published literature, it is clear that QoS enforcement 
features are conspicuous by their absence
in general-purpose cloud computing resource management surveys, 
see \tablename~\ref{tab:prev_surveys}. 

\begin{table}[htbp]
    \centering
    \caption{Previous surveys on shared resource management.}
    \label{tab:prev_surveys}
    \small
    \begin{tabular}{lcccc}
        \toprule
        \multirow{2}{*}{\textbf{Ref.}} & \multirow{2}{*}{\textbf{QoS EF*}} & \multicolumn{3}{c}{\textbf{Discussion Context}} \\ 
        \cline{3-5} 
                      & &  \textbf{RT*} & \textbf{VNFs*} & \textbf{Cloud} \\ 
                      \midrule
                      
         \cite{mittal_survey_2017} & \checkmark & & & \\
         \cite{gracioli_survey_2015, lugo_survey_2022} & \checkmark & \checkmark & &  \\ 
         \cite{kawashima_vision_2021} & \checkmark & & \checkmark &  \\
         \cite{hameed_survey_2016, liaqat_federated_2017, hong_resource_2019, 
         bermejo_virtual_2020, helali_survey_2021, chaurasia_comprehensive_2021, 
         carrion_kubernetes_2022, huang_survey_2023} & & & & \checkmark \\ 
         \midrule
         This Survey & \pmb{\checkmark} & & & \pmb{\checkmark} \\

         \bottomrule
    \end{tabular}

    \vspace{.25em}
    \scriptsize{
    *Abbreviations --- QoS EF: QoS Enforcement Feature Coverage, \\ RT: Real-time, VNFs: Virtual Network Functions.
    }
    
\end{table}

Cloud environments 
rarely see hard execution time deadlines on tasks, as is common in real-time systems, 
but rather statements of performance intent in the form of service level objectives (SLOs). 
Resource management optimization for cloud environments is therefore typically
discussed in a broader context and at a higher level of abstraction than real-time systems.
Thus, lower-level sources of congestion such as LLC and memory bandwidth are often 
overlooked, excluded, or disregarded in cloud resource 
management surveys~\cite{hameed_survey_2016,liaqat_federated_2017,hong_resource_2019,
bermejo_virtual_2020,helali_survey_2021,chaurasia_comprehensive_2021,carrion_kubernetes_2022,
huang_survey_2023}.
Even so, increased resource efficiency and task consolidation achieved 
through QoS enforcement feature utilization is not exclusive to real-time systems.

\revised{We thus recognize the need for a broad review of literature approaching shared 
resource management using hardware-level QoS enforcement features to gain insight
into past and current trends in research.
This review serves as a basis for understanding how the academic community approaches
resource management issues using hardware-implemented QoS enforcement features.
By understanding the current approaches and their traits, we aim to detail
the major challenges that hinder the practicality and widespread adoption of QoS 
enforcement in the cloud.}

\subsection{Related Surveys}

A few systematic reviews have previously been conducted that cover neighboring
and overlapping areas of research.
In their 2015 article, \citeauthor*{gracioli_survey_2015}~\cite{
gracioli_survey_2015} surveyed cache management approaches for embedded real-time systems. 
They recognize that server processors supporting Intel CAT, the only commercially available
feature at the time, are not targeted at the embedded systems market surveyed in their paper. 
However, they still make the case for cache allocation's usefulness for the optimization
of soft real-time applications.

Publications with a focus on resource management in the cloud often approach the problem at a
higher abstraction layer than papers with lower-level performance as their context, such 
as those from the real-time computing community. This often means that CPU cache and memory 
bandwidth management on the node level is out of scope. Such is the case in the surveys by
\citeauthor*{hameed_survey_2016}~\cite{hameed_survey_2016}, \citeauthor*{liaqat_federated_2017}~\cite{
liaqat_federated_2017}, and \citeauthor*{hong_resource_2019}~\cite{hong_resource_2019}.

With a clear focus on cache partitioning, \citeauthor*{mittal_survey_2017}~\cite{mittal_survey_2017}
surveyed techniques for mitigating the increased shared cache congestion that follows from increasing 
core counts in processors. This includes early publications utilizing CAT, as well as a large number of
software-based approaches using techniques such as page coloring.

\citeauthor*{bermejo_virtual_2020}~\cite{bermejo_virtual_2020},
\citeauthor*{helali_survey_2021}~\cite{helali_survey_2021}, and 
\citeauthor*{chaurasia_comprehensive_2021}~\cite{chaurasia_comprehensive_2021} surveyed techniques of
workload consolidation in the cloud. The first with a specific focus on virtual machine (VM) 
consolidation. \citeauthor*{kawashima_vision_2021}~\cite{kawashima_vision_2021} acknowledge the
usefulness of cache partitioning when optimizing the performance of virtual network functions (VNFs)
that power much of the networking in modern clouds.

\citeauthor*{lugo_survey_2022}~\cite{lugo_survey_2022} 
surveyed works on process interference for real-time systems published between 2015 and 2020. 
The works included proposed techniques that reduce congestion in LLC, 
main memory, as well as in the main memory bus. A few of the works utilize QoS enforcement 
features, while others use software-based approaches. 

With the rise of Kubernetes as the de-facto standard container orchestrator, 
\citeauthor*{carrion_kubernetes_2022}~\cite{carrion_kubernetes_2022} identifies the need 
to survey the current
approaches and ongoing issues in Kubernetes scheduling. They find that Kubernetes is used in
many different environments, running different types of workloads on different types of hardware, 
which dictates the type of scheduling approaches leveraged.

In a recent publication, \citeauthor*{huang_survey_2023}~\cite{huang_survey_2023} review works on resource 
management in the context of cloud-native mobile computing. This has received an increasing 
amount of attention recently due to the architecture of fifth and sixth-generation mobile communication networks 
(5G and 6G) being based on cloud-native technology. Most of the works reviewed focus on the management of networking 
resources through the utilization of network slicing or network virtualization.

\section{Review Method} \label{sec:method}

This section outlines the systematic approach used when compiling publications leveraging 
hardware QoS enforcement features. 
We do this to ensure an unbiased and reproducible selection of publications for inclusion 
in our review. 

\subsection{Search Strategy}

\begin{figure}[htbp]
    \centering
    \small
    \includegraphics{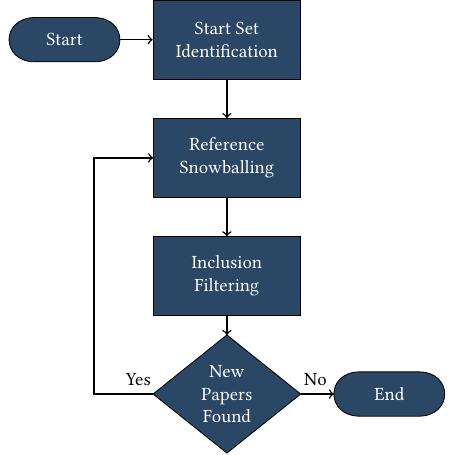}
    \caption{Reference snowballing search strategy used.}
    \label{fig:snowball}
    \Description{Reference snowballing search strategy used.}
\end{figure}

Our search strategy is based on the reference snowballing process outlined by 
\citeauthor{wohlin_guidelines_2014}~\cite{wohlin_guidelines_2014} and follows a
simple algorithm. First, a starting set of core publications is identified by 
querying a digital literature database using a defined set of search terms and inclusion
criteria. Then, both backward and forward snowballing is used to identify 
additional papers that fit the inclusion criteria. The snowballing process is 
repeated after identified papers are added to the set of included papers.
This process is repeated until no additional papers that 
fit the inclusion criteria are found. This process is outlined in 
\figurename~\ref{fig:snowball}.

Backward snowballing is the process of observing the reference lists of 
included papers. These lists are scanned and filtered to identify additional 
papers that meet the inclusion criteria. Similarly, forward snowballing identifies
papers that reference already included papers. Any publication that fits the 
inclusion criteria is added to the set of included papers. The snowballing
search space may grow every iteration that one or more papers are added. Once no
new paper that fits the inclusion criteria is found during a snowballing iteration,
the search is stopped as the set of potential papers no longer grows.

Our motivations for selecting this approach
instead of a more traditional database search approach (i.e. 
\citeauthor*{kitchenham_evidence-based_2004}~\cite{kitchenham_evidence-based_2004}) are as follows: 
We survey how the academic community leverages a defined set of hardware 
features in research. We can thus find our starting set of articles by
explicitly searching for the names of the features. This should prove a good 
basis for snowballing as authors typically 
cite a small set of early papers as sources for further details of a feature. Such 
publications become natural ``roots'' in our reference tree.
Moreover, it was shown by 
\citeauthor*{badampudi_experiences_2015}~\cite{badampudi_experiences_2015} that 
snowballing is comparable, if not more reliable, than a traditional database search 
approach.

\subsection{Threats to Review Validity} \label{sec:validity}

The snowballing process is highly dependent on the quality of the starting set of papers
used to seed the search. A starting set that is too small, or not diverse enough 
can leave ``dark spots'' where many papers are left out. One reason for this is a same 
author bias that may emerge. As researchers often cite their previous work due 
to relevance, a snowballing search may become heavily biased towards a 
few researchers' works~\cite{jalali_systematic_2012}. The same phenomenon may be true
for publishers instead of individual researchers~\cite{wohlin_guidelines_2014}.
\revised{However, the nature of our review and its focus on a select set of hardware
features should limit the impact of this bias, as researchers reference a set of
early papers when discussing these features.}

It is also possible that our review is incomplete as researchers may be using hardware QoS 
features without explicitly referencing a source for that feature. This would mean that unless
the paper is cited by another included paper, it would not be found. There is the possibility
that it could be found in the search for the starter set of papers, however, this is unlikely.

\subsection{Inclusion Criteria}

The criteria used to filter publications are derived from our review objectives. The inclusion 
criteria used are as follows:

\begin{enumerate}
    \item The proposal utilizes one or more hardware-based QoS enforcement features available on 
        commercial off-the-shelf x86 chips. 
    \item The research uses the feature or features first-hand, not through re-implementation of
        a previous proposal.
    \item The publication is four pages or longer.
    \item \label{cond:date} The paper was published before July 1st, 2023.
    \item The paper is written in English.
\end{enumerate}

No starting date of the publication time span is defined as hardware
QoS features have only recently become available in commodity off-the-shelf server processors. 
Intel's CAT was originally introduced in the Xeon E5-2600 v3 family of 
processors~\cite{herdrich_cache_2016} which launched in the third quarter of 2014. 
Thus, 2014 serves as a starting date for CAT publications.

\revised{We focus our review on hardware-based control mechanisms, despite the 
long-standing existence of software-based QoS enforcement approaches. 
Software-based methods, such as page coloring~\cite{taylor_tlb_1990} or MemGuard~\cite{yun_memguard_2013}, 
offer similar functionalities to their hardware counterparts 
and have likely influenced the design of current hardware features.}

\revised{While processor-agnostic software control strategies have their own 
advantages and use cases, potentially addressing some limitations 
of current hardware mechanisms, chip manufacturers continue to 
enhance hardware-level control mechanisms with each new generation~\cite{sohal_closer_2022}. 
This ongoing improvement suggests that the future of QoS enforcement lies 
in hardware-level mechanisms, which require lower-level control and 
platform-specific optimizations for optimal performance. Therefore, we 
concentrate on hardware-level control mechanisms in this review.}

\subsection{Starting Set Identification}

Google Scholar\footnote{\url{http://scholar.google.com}} was used as a search engine when querying
for potential starting set articles. This had the advantage of minimizing potential publisher
bias as the search covers a large number of publishing sources.
A separate search was conducted for each of the hardware
features of interest. The exact search terms used were: 
\begin{enumerate}
    \item \label{search:cat} ``intel cache allocation technology''
    \item \label{search:mba} ``intel memory bandwidth allocation''
    \item \label{search:qose} ``amd64 technology platform quality of service extensions''
\end{enumerate}

\begin{figure}[htbp] 
\begin{multicols}{2}
    \begin{table}[H]
        \centering
        \small
        \caption{Start set of publications used to seed the snowballing search ordered by
        date of publication from oldest to most recent.}
        \label{tab:startset}
        \begin{tabular}{ccl}
            \toprule
            \textbf{Ref.} & \textbf{Search} & \textbf{Authors} \\ 
            \midrule
            
            \cite{funaro_ginseng_2016} & \ref{search:cat} & \citeauthor{funaro_ginseng_2016} \\
            \cite{xu_vcat_2017} & \ref{search:cat} & \citeauthor{xu_vcat_2017} \\
            \cite{selfa_application_2017} & \ref{search:cat} & \citeauthor{selfa_application_2017} \\
            \cite{xu_dcat_2018} & \ref{search:cat} & \citeauthor{xu_dcat_2018} \\
            \cite{pons_improving_2018} & \ref{search:cat} & \citeauthor{pons_improving_2018} \\
            \cite{park_hypart_2018} & \ref{search:mba} & \citeauthor{park_hypart_2018} \\
            \cite{xiang_dcaps_2018} & \ref{search:cat} & \citeauthor{xiang_dcaps_2018} \\
            \cite{kim_application_2019} & \ref{search:cat} & \citeauthor{kim_application_2019} \\
            \cite{park_copart_2019} & \ref{search:mba} & \citeauthor{park_copart_2019} \\
            \cite{xiang_emba_2019} & \ref{search:mba} & \citeauthor{xiang_emba_2019} \\
            \cite{zhang_libra_2021} & \ref{search:qose} & \citeauthor{zhang_libra_2021} \\
            \cite{sohal_closer_2022} & \ref{search:cat} & \citeauthor{sohal_closer_2022} \\
            \cite{farina_assessing_2022} & \ref{search:mba} & \citeauthor{farina_assessing_2022} \\
            
            \bottomrule
        \end{tabular}
    \end{table}
   
   \columnbreak
   
    \begin{figure}[H]
        \centering
        \small
        \footnotesize
        \includeinkscape[angle=270]{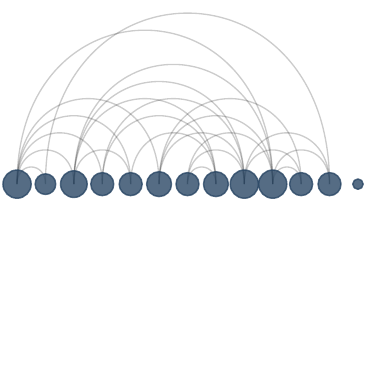}
        \caption{Arc diagram illustrating the reference relationships between the papers 
            selected for the starter set. Publications are chronologically ordered 
            from top to bottom. The size of the nodes corresponds to the number of
            connections they have.}
        \label{fig:start_tree}

        \Description{Arc diagram illustrating the reference relationships between the papers 
            selected for the starter set. Publications are chronologically ordered 
            from top to bottom. The size of the nodes corresponds to the number of
            connections they have.}
    \end{figure}
\end{multicols}
\end{figure}

A set of seed publications were selected from these lists of publications based on our inclusion criteria,
publisher diversity, and relevance. This start set of
papers are listed in \tablename~\ref{tab:startset} and \revised{a graph of their 
reference relationships} may be seen in
\figurename~\ref{fig:start_tree}. \revised{While this starting set is small and
may contain important gaps in the literature, it is a large and diverse enough
starting point. The snowballing process will help fill in these gaps as the 
included papers reference, or are referenced by, other important publications.}

\subsection{Reference Snowballing}

\begin{figure}[htbp]
    \centering
    \small
    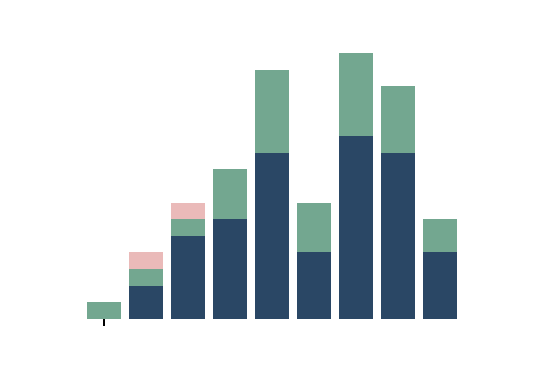
    \caption{Distribution of articles utilizing platform features grouped by their year of publication. Papers
    are classified as academic when all authors are affiliated with academic institutions, whereas cooperative
    publications have one or more authors affiliated with a private company. Industry papers have no affiliation
    with an academic institution. (*2023 is limited to the first six months of the year.)}
    \label{fig:hardware-year}
    \Description{Distribution of articles utilizing platform features grouped by their year of publication. Papers
    are classified as academic when all authors are affiliated with academic institutions, whereas cooperative
    publications have one or more authors affiliated with a private company. Industry papers have no affiliation
    with an academic institution. (*2023 is limited to the first six months of the year.)}
\end{figure}

Using the snowballing search method, we iterated upon the defined starting set.
At the time of the final snowballing algorithm iteration, the search set had increased to $\searchsize$
references. Out of these $\searchsize$ publications, $\surveysize$ papers were 
identified to meet the inclusion criteria of our review. 
\figurename~\ref{fig:hardware-year} illustrates the included articles' publication dates and how interest
from researchers in both academia and industry has evolved over time.

It is notable that if reproduced, the search would yield a larger search set than found in this
survey. This is due to the included articles being cited as sources in new papers published
over time. However, due to the publication date limitation (inclusion criteria~(\ref{cond:date})),
the set of included articles is deterministic and a reproduction will produce the same set
of included articles.

\section{Taxonomy} \label{sec:taxonomy}

\begin{figure}[htbp]
    \centering
    \small
    \includegraphics{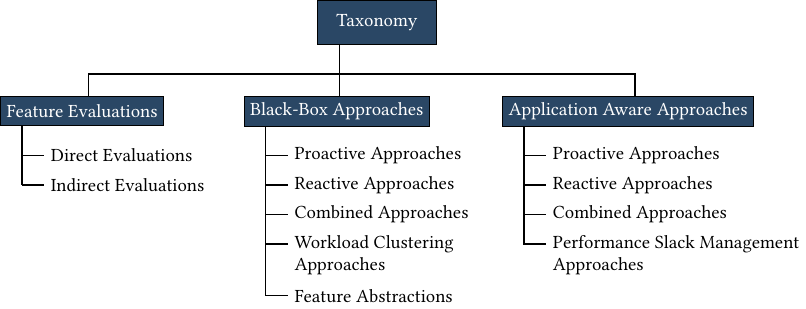}
    \caption{Taxonomy organization.}
    \label{fig:hardware-layout}
    \Description{Taxonomy organization.}

\end{figure}

We group included publications into three broad categories:
\begin{itemize}
    \item \emph{Feature evaluations} are works that showcase and evaluate the hardware mechanisms
       from different points of view, in many cases through empirical study.  
    \item \emph{Black-box approaches} propose methods or controllers that only observe metrics from 
        hardware or the operating system, treating software as opaque ``black boxes''. 
    \item \emph{Application aware approaches} observe metrics that are specific to 
        the executing software\footnote{E.g. application latencies, throughput, etc.} 
        in addition to hardware metrics. 
\end{itemize}
The layout of this section is based on this classification 
and may be observed in \figurename~\ref{fig:hardware-layout}.

\subsection{Feature Evaluations}

The function of QoS enforcement features implemented in hardware has been experimentally
analyzed by a number of different 
works~\cite{sohal_closer_2022, zhang_libra_2021, herdrich_cache_2016, farina_assessing_2022,
ihre_sherif_evaluation_2021,schramm_impact_2019,kim_simple_2017,veitch_performance_2017,castiglione_optimal_2018}. 
It is important to verify the behavior of hardware features so that they may be used as
a basis for other research. However, the functionality may change between chip generations 
due to architectural differences~\cite{sohal_closer_2022}. This further highlights the importance
of analyzing the behavior of new implementations as the features mature over time. 
Additionally, some works~\cite{bechtel_denial--service_2022, qiu_is_2021, farshin_reexamining_2020} 
evaluate if QoS enforcement features may be used as solutions to problems in other research areas.
Other publications~\cite{shang_coplace_2021} use QoS enforcement features to demonstrate problems and
subsequent solutions to resource management problems.

The publications evaluating hardware features are compiled in \tablename~\ref{tbl:feature-evaluations}.
Notably, all of the studies classified as feature evaluations have evaluated and 
verified the functionality of Intel implementations, AMD QoSE has not been 
functionally evaluated in the current literature.

\begin{table*}[t]
    \centering
    \small
    \caption{Publications presenting evaluations of hardware features.}
    \label{tbl:feature-evaluations}
    \begin{tabular}{cp{.45\textwidth}ccc}
        \toprule
        \multirow{2}{*}{\textbf{Ref.}} & \multirow{2}{*}{\textbf{Evaluation}} & \multicolumn{3}{c}{\textbf{Feature(s)}} \\ 
        \cline{3-5} 
                      & &  \textbf{Intel CAT} & \textbf{Intel MBA} & \textbf{AMD QoSE} \\ 
        \midrule \multicolumn{5}{c}{Direct Evaluations} \\ \midrule

        \cite{herdrich_cache_2016} & Introduction and feature showcase of CAT. & \checkmark & & \\
        \cite{sohal_closer_2022} & Experimental evaluation of CAT and MBA. & \checkmark & \checkmark & \\
        \cite{zhang_libra_2021} & Highlights the shortcomings of MBA. & \checkmark & \checkmark & \\
        \cite{farina_assessing_2022} & Experimental evaluation of MBA for RT systems. & & \checkmark &  \\
        \cite{ihre_sherif_evaluation_2021} & Experimental evaluation of CAT. & \checkmark & & \\
        \cite{schramm_impact_2019} & Experimental evaluation of CAT for VNF systems. & \checkmark & & \\
        \cite{qin_when_2018} & Study of cache partitioning in correlation to workload cache characteristics. & \checkmark & & \\
        \cite{kim_simple_2017} & Evaluation of cache partitioning as a tool to meet workload latency targets. & \checkmark & & \\
        \cite{veitch_performance_2017} & Experimental evaluation of CAT for VNF systems. & \checkmark & & \\
        \cite{castiglione_optimal_2018} & Highlights the potential usefulness of CAT for security purposes. & \checkmark & & \\

        \midrule \multicolumn{5}{c}{Indirect Evaluations} \\ \midrule
        
        \cite{qiu_is_2021} & Evaluation of FaaS for LC workload hosting. & \checkmark & \checkmark & \\
        \cite{bechtel_denial--service_2022} & Evaluation of LLC partitioning as DoS protection on iGPU platforms. & \checkmark & & \\
        \cite{farshin_reexamining_2020} & Evaluation of Intel Data Direct I/O. & \checkmark & \checkmark & \\
        \cite{shang_coplace_2021} & Intelligent Guest to Host Page Mapping. & \checkmark &  & \\

        \bottomrule
    \end{tabular}
\end{table*}

\subsubsection{Direct Evaluations}

As a way of introducing Intel CAT to the scientific community, 
\citeauthor*{herdrich_cache_2016}~\cite{herdrich_cache_2016} present a feature showcase
in their 2016 paper. The publication addresses the specifics of the 
implementation and highlights the importance of the feature by suggesting use cases 
for the new technology. To back up their claims, the effect on the performance
of workloads constrained by LLC allocations is presented and discussed.

In their recent study, \citeauthor*{sohal_closer_2022}~\cite{sohal_closer_2022} conduct
an experimental evaluation of Intel's CAT and MBA. It is found that the behavior of the
features differs between chip generations. Therefore, they suggest using their evaluation 
approach for coming architectural versions in order to verify the behavior of such versions.

\citeauthor*{zhang_libra_2021}~\cite{zhang_libra_2021} conducts a smaller study that
highlights the shortcomings of Intel MBA. This evaluation is used as motivation
when introducing their own software-based approach to memory bandwidth allocation.

Through an experimental analysis, \citeauthor*{farina_assessing_2022}~\cite{farina_assessing_2022}
evaluate the effectiveness and behavior of MBA for resource limitation in real-time systems. They
conclude that the hardware-based approach to memory bandwidth limiting outperforms
similar software-based approaches such as MemGuard~\cite{yun_memguard_2013}.

In their master thesis, \citeauthor*{ihre_sherif_evaluation_2021}~\cite{ihre_sherif_evaluation_2021} 
evaluate the performance 
implications of LLC partitioning on three commonly used workloads: bzip2, redis, and Graph500.
They find that tuned partitioning may increase the performance of the aforementioned workloads
by up to $5.8\%$, $8.6\%$, and $12.0\%$, respectively, compared to an unmanaged system
when co-located with a noisy neighbor.

\citeauthor*{schramm_impact_2019}~\cite{schramm_impact_2019} explore the effectiveness of CAT
as a method of improving the performance of virtual network functions (VNFs). They find that
sufficient LLC allocation may increase the throughput of a VNF by up to $33 \%$ during heavy
load scenarios.

To better understand the implications of cache partitioning on workloads with
different characteristics, \citeauthor*{qin_when_2018}~\cite{qin_when_2018}, 
\citeauthor*{kim_simple_2017}~\cite{kim_simple_2017}, and 
\citeauthor*{veitch_performance_2017}~\cite{veitch_performance_2017} conduct empirical studies 
where the behavior of a range of workloads under different cache partitioning schemes is examined.
They find that the tuning of cache allocation schemes is generally beneficial to ensuring
predictable performance of workloads.

While not strictly presenting a feature evaluation, \citeauthor*{castiglione_optimal_2018} discuss
the potential usefulness of cache allocation in preventing and mitigating cache side-channel 
attacks~\cite{castiglione_optimal_2018}. 
They observe the behavior enabling the side-channel attack vector and argue for the applicability
of cache partitioning as a possible solution to the problem. However, determining if cache allocation upholds
the promises is left as future work.

\subsubsection{Indirect Evaluations}

With the rise of serverless computing and function-as-a-service (FaaS) platforms, an increasing
number of practitioners are looking to the promises of ease-of-use and pay-per-use of the
paradigm. \citeauthor*{qiu_is_2021}~\cite{qiu_is_2021} recognize the need to evaluate
the feasibility of FaaS as a model to host LC workloads. In their evaluation, they 
include LLC and memory bandwidth partitioning as potential ways of ensuring consistent performance.
Their findings conclude that FaaS is not yet ready for LC workloads and that the aforementioned
partitioning techniques would have to support orders of magnitude more CLOSs before  
QoS of FaaS functions can be sufficiently enforced.

On platforms with integrated graphics, LLC is often shared between the CPU and GPU. 
This becomes an attack vector for denial-of-service (DoS) attacks. 
\citeauthor*{bechtel_denial--service_2022}~\cite{bechtel_denial--service_2022} evaluates if LLC 
partitioning using CAT provides sufficient protection to mitigate such attacks. They conclude
that CAT may indeed provide spatial isolation, but that there are still cases where temporal
isolation can not be achieved. This in turn leads to sophisticated attacks still being viable.

To better understand the mechanics of Intel's Data Direct I/O (DDIO) 
and its correlation to LLC cache ways, 
\citeauthor*{farshin_reexamining_2020}~\cite{farshin_reexamining_2020} conducts an empirical 
study to gain a better understanding of the inner workings of the technology.
With the knowledge gained from this study, they discuss its effectiveness and 
shortcomings. General advice regarding how DDIO should be utilized depending 
on the use case is also presented.

\citeauthor*{shang_coplace_2021}~\cite{shang_coplace_2021} utilize LLC partitioning to
demonstrate that caching inefficiencies exist in the way hosts map guest VM memory pages to 
physical memory. Their proposed solution, \emph{CoPlace} mitigates this problem and reduces
cache conflict misses that exist even when a guest is given exclusive access to some subset 
of LLC.

\subsubsection{Summary}

The hardware features of interest have been evaluated in many works from different
points of view. Some with the intention of showcasing and introducing them to the scientific 
community, and some to see if they can help solve a problem in another research field. It is 
clear that the behavior may change between chip architecture generations, especially
in more recent and less mature implementations. Behavior should thus be validated in
future generations as the implementations mature.

\subsection{Black-Box Approaches} \label{sec:black-box}

\begin{figure}[htbp]
    \centering
    \small
    \includegraphics{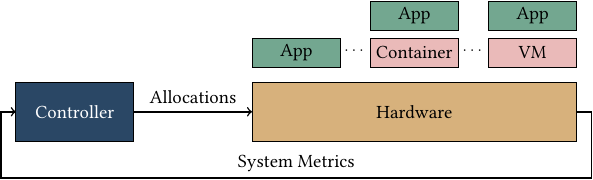}
    \caption{\revised{Illustration of black-box approaches and their reliance on system and hardware-only metrics. 
    Such metrics may include cache miss rates, memory bandwidth usage, and instructions per cycle numbers, for example.
    System metrics are generic and exist for all workloads.}}
    \label{fig:black-box}
    \Description{Illustration of black-box approaches and their reliance on system metrics.}
\end{figure}

The ability to tune a system without knowledge of the applications it is hosting can be useful. 
However, doing so may be challenging considering that different applications are sensitive to different 
bottlenecks~\cite{cortez_resource_2017}. Even so, it may be required in some cases due to 
privacy concerns. Such is the case with public cloud providers that are unable to observe the 
workloads running in customers' virtual machines~\cite{shahrad_provisioning_2021}. 
Consequently, tuning must be done by observation of system metrics only. 
Thus, tuning becomes workload agnostic, as all processes are observed using
the same system and operating system metrics. \revised{This is illustrated in \figurename~\ref{fig:black-box}.}

Proposals leveraging the surveyed platform features that treat workloads as ``black boxes'' are
presented in \tablename~\ref{tbl:black-box}. Black box approaches are split into subclasses: 
Proactive approaches, reactive approaches, combined approaches, workload clustering approaches, and feature abstractions.

\begin{table}[htbp]
    \centering
    \small
    \caption{Publications leveraging only hardware and operating system-level metrics.}
    \label{tbl:black-box}
    \begin{tabular}{ccclll}
        \toprule
        \textbf{Ref.} & \textbf{CA*} & \textbf{MBL*} & \textbf{DC*} 
            & \textbf{Optimization} & \textbf{Proposed Technique} \\
        \midrule \multicolumn{6}{c}{Proactive Approaches} \\ \midrule
        
        \cite{xiao_cppf_2019} & \checkmark &  & Generic
            & Performance & Prefetch-Aware Heuristic LLC Partitioning \\
        \cite{kim_application_2019} & \checkmark &  & Generic
            & Perf. Prediction & ML Performance Prediction Model \\
        \cite{aupy_co-scheduling_2019} & \checkmark &  & HPC
            & Perf. Prediction & Mathematical Modelling of Execution Times \\
        \cite{durner_towards_2019} & \checkmark & & VNFs
            & Utilization & Static Heuristic Optimization Algorithm \\
        \cite{funaro_ginseng_2016} & \checkmark & & Cloud
            & Performance & Market-Driven Auction-Based Scheme \\
        \cite{shahrad_provisioning_2021} & \checkmark & & Cloud  
            & Performance & Cache-Aware VM Scheduler \\
        \cite{liu_catalyst_2016} & \checkmark & & Security
            & Security & LLC Side-channel Attack Protection Scheme \\
        \cite{zeng_playing_2021} & \checkmark & & Generic
            & Performance & Greedy Static LLC Allocation Scheme \\
        \cite{chung_enforcing_2019} & \checkmark & \checkmark & Generic
            & Performance & LLC-Aware Memory Bandwidth Management \\ 
        \cite{thomas_dark_2018} & \checkmark & & VNFs
            & Performance & Memory Hierarchy Management Framework \\
        \cite{sprabery_novel_2017} & \checkmark & & Security
            & Security & LLC Side-Channel Attack Protection Scheme \\
        
    \midrule \multicolumn{6}{c}{Reactive Approaches} \\ \midrule
    
        \cite{navarro-torres_balancer_2023} & \checkmark & \checkmark & Generic
            & Perf., Fairness & Heuristic LLC and MB Management Scheme \\ 
        \cite{park_hypart_2018} & & \checkmark & Generic
            & Performance & Explorative MB Partitioning Algorithm \\
        \cite{park_copart_2019} & \checkmark & \checkmark & Generic
            & Perf., Fairness & Explorative LLC and MB Partitioning \\
        \cite{xu_dcat_2018} & \checkmark &  & Generic
            & Performance & Heuristic Program Phase Detection \\
        \cite{pons_cache-poll_2022} & \checkmark & & Generic
            & Perf., Fairness & Heuristic LLC Partitioning Scheme \\
        \cite{farina_enabling_2023} & & \checkmark & RT
            & Performance & Heuristic Bandwidth Guarantees \\
        \cite{roy_satori_2021} & \checkmark & \checkmark & Generic
            & Perf., Fairness & Bayesian Optimization Based Algorithm \\
        \cite{chen_alita_2020} & \checkmark & \checkmark & Cloud
            & Fairness & Heuristic VM Resource Fairness Algorithm \\
        \cite{xiang_dcaps_2018} & \checkmark &  & Generic
            & Performance & Simulated Annealing Optimization \\
        \cite{yuan_dont_2021} & \checkmark & & Generic
            & Performance & Heuristic, DDIO Aware LLC Management \\
        \cite{yi_mt2_2022} & & \checkmark & Generic
            & Performance & Heuristic MB Management Algorithm \\
        \cite{yao_leveraging_2019} & \checkmark & & Security 
            & Security & Cache Timing Channel Attack Mitigation \\
            
    \midrule \multicolumn{6}{c}{Combined Approaches} \\ \midrule
    
        \cite{li_pfa_2022} & \checkmark & & Generic
            & Perf., Fairness & Heuristic LLC Management Algorithm \\
        \cite{garcia-garcia_lfoc_2019} & \checkmark &  & Generic
            & Fairness & Heuristic Cache-Clustering Algorithm \\
        \cite{saez_lfoc_2022} & \checkmark & & Generic
            & Fairness & Heuristic Cache-Clustering Algorithm \\
        \cite{pons_improving_2018} & \checkmark &  & Generic
            & Perf., Fairness & Heuristic LLC Partitioning Scheme \\
        \cite{pons_phase-aware_2020} & \checkmark &  & Generic
            & Perf., Fairness & Heuristic LLC Partitioning Scheme \\
        \cite{chen_drlpart_2021} & \checkmark & \checkmark & Generic 
            & Performance & Deep Reinforcement Learning Model \\
        \cite{chatterjee_com-cas_2023} & \checkmark & & Generic 
            & Performance & Compiler Guided LLC Management \\
        \cite{zhang_zeus_2021} & \checkmark & & Cloud
            & Utilization & Kubernetes Resource Management Scheme \\
        \cite{zhu_kelp_2019} & \checkmark & & ML
            & Performance & Data Management for ML Systems \\
            
    \midrule \multicolumn{6}{c}{Workload Clustering Approaches} \\ \midrule
    
        \cite{tang_themis_2023} & \checkmark & \checkmark & Generic
            & Perf., Fairness & K-Means Classification of Guest VMs \\
        \cite{gifford_dna_2021} & \checkmark & & Generic
            & Performance & Workload Clustering Technique \\
        \cite{xiang_emba_2019} & & \checkmark & Generic
            & Performance & Heuristic Hierarchical Clustering Algorithm \\
        \cite{selfa_application_2017} & \checkmark &  & Generic
            & Fairness & K-Means Classification of Workloads \\
        \cite{qiu_machine-learning-based_2022} & \checkmark & & Cloud
            & Performance & SVM Classification and BO LLC Partitioning \\
            
    \midrule \multicolumn{6}{c}{Feature Abstractions} \\ \midrule
        \cite{xu_vcat_2017} & \checkmark & & VMs
            & Performance & Virtualization of CAT \\

    \bottomrule
    
    \multicolumn{6}{c}{\scriptsize *Abbreviations --- CA: Cache Allocation, MBL: Memory Bandwidth Limiting, DC: Discussion Context.} 

    \end{tabular}
\end{table}

\begin{figure}[htbp]
    \centering
    \begin{subfigure}[b]{0.38\textwidth}
        \centering
        \includegraphics{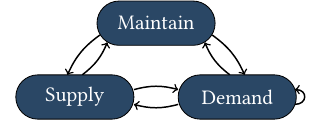}
        \caption{\revised{CoPart resource FSM illustration.}}
        \label{fig:copart-fsm}
        \Description{CoPart resource FSM illustration.}
    \end{subfigure}
    \begin{subfigure}[b]{0.61\textwidth}
        \centering
        \includegraphics{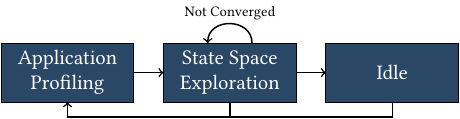}
        \caption{\revised{CoPart controller execution flow.}}
        \label{fig:copart-flow}
        \Description{CoPart controller execution flow.}
    \end{subfigure}
    \caption{\revised{Illustration of core components of CoPart by \citeauthor{park_copart_2019}.}}
    \label{fig:copart}
    \Description{}
\end{figure}

\revised{As an example of an important black-box approach, \emph{CoPart}~\cite{park_copart_2019} aims 
to enhance both performance and fairness of system workloads by classifying 
them as either sensitive or insensitive to cache and memory bandwidth based 
on profiling. This classification enables the controller to construct two 
finite state machines (FSMs) for each application, representing its cache and 
memory bandwidth characteristics, respectively (see Figure \ref{fig:copart-fsm}).
Each FSM contains three states: Supply, where the application can have its allocation reduced without significant performance impact;
Demand, where performance would improve if allocated additional resources;
And maintain, if performance would degrade if the allocation is reduced, but would only marginally improve if increased.
Using these FSMs, CoPart explores the available states to achieve an optimized 
resource allocation scheme based on current system metrics and by mapping 
against the Hospitals/Residents (HR) problem~\cite{gale_college_1962}. The execution flow of CoPart 
is illustrated in Figure \ref{fig:copart-flow}.}

\subsubsection{Proactive Approaches} 

\revised{Publications featuring approaches classified as proactive attempt to make 
resource management decisions in advance. They may enforce static partitioning 
schemes by profiling workloads and creating a fixed resource division or try to 
forecast future workload resource requirements based on past and current metrics.}

By profiling workloads as sensitive, or non-sensitive, to hardware prefetching of
memory, \citeauthor*{xiao_cppf_2019}~\cite{xiao_cppf_2019} propose a controller that
aims to improve LLC management on modern systems equipped with hardware prefetching.
They call their approach \emph{CP$_{pf}$}, which partitions LLC in accordance with the
determined workload classifications. The system allocates a single exclusive
cache way to each prefetching sensitive workload, leaving the rest to be shared among 
non-prefetching sensitive workloads.

With the goal of predicting how a process' instructions per cycle (IPC) metric changes
as a result of a change in its available LLC, 
\citeauthor*{kim_application_2019}~\cite{kim_application_2019} propose \emph{POCAT}. 
POCAT is a trained machine learning model that can predict changes to IPC with an error 
of 4.7\% without affecting current LLC allocations.

To get a better understanding of how runtimes of BE high-performance computing (HPC) workloads 
are affected by core and LLC allocations, 
\citeauthor*{aupy_co-scheduling_2019}~\cite{aupy_co-scheduling_2019} create 
a mathematical model that predicts how job execution times are affected by 
changes in scarce shared resource availability. Using their model, they are able to more efficiently partition 
resources among workloads to increase the efficiency of a system.

A simple heuristic algorithm to partition LLC among VNFs is proposed by 
\citeauthor*{durner_towards_2019}~\cite{durner_towards_2019}. Their algorithm
distributes available cache ways based on CPU utilization, which results in
lowered total system utilization, increasing energy efficiency and QoS.
However, only static configurations are considered
where virtual network functions (VNFs) and their loads do not change. This makes their algorithm
valid for very specific use cases only.

In an attempt to make LLC partitioning more effective, \emph{Ginseng}~\cite{funaro_ginseng_2016}
allow consumers (i.e. workloads) to bid on LLC space using Vickrey-Clarke-Groves auctioning. This allocates LLC
to the process that would see the largest economic benefit from having said space. As the auctioning
and allocation cycle is kept short, processes may acquire more LLC space at critical phases in 
its operation, while seceding it at times of less sensitivity.

Public cloud providers aim to reduce their total cost of hardware ownership by 
optimizing its utilization. Making a VM scheduler LLC aware allows the provider to 
give LLC QoS guarantees to their customers as LLC ways may be exclusively, or partially
exclusively, allocated to specific VMs. This is the goal of \emph{CacheSlicer}, 
a VM scheduler proposed by
\citeauthor*{shahrad_provisioning_2021}~\cite{shahrad_provisioning_2021} 
that schedules VMs onto nodes based on low-cost, normal, and cache-sensitive VM categories. 

In public clouds, where different clients' VMs share physical hosts, malicious actors
may compromise security by extracting sensitive information from a co-located guest 
by manipulation of the LLC. \emph{CATalyst}, as proposed by 
\citeauthor*{liu_catalyst_2016}~\cite{liu_catalyst_2016}, is a module that integrates
into hypervisors that protect against such LLC side-channel attacks by partitioning
the LLC to provide security guarantees.

\citeauthor{zeng_playing_2021} find that a simple static allocation of LLC
is not sufficient to improve system performance when software prefetching of memory
is used~\cite{zeng_playing_2021}. They show that improvements can be made 
even with a simple greedy partitioning scheme.

An observation that allocating more LLC to an LC application may actually bring worse
performance is made by \citeauthor*{chung_enforcing_2019}~\cite{chung_enforcing_2019}.
This is due to the memory controller being overwhelmed by the requests of the remaining
workloads that miss LLC more often due to less remaining usable after partitioning. The
paper details their approach to mitigate this issue and the ``memory virtual channels''
they use to prioritize and order memory requests such that the LC application is
guaranteed prioritized access to main memory as well as LLC.

\citeauthor*{thomas_dark_2018} observe that as networks approach transfer speeds of 100 gigabits per second
and beyond, the memory system becomes a bottleneck that can a cause large number of packet drops 
if incorrectly utilized~\cite{thomas_dark_2018}. As an attempt to mitigate this issue, they
propose \emph{CacheBuilder}. Cache builder is a system that, among other techniques, utilizes cache
allocation to provide applications with limited control of the memory hierarchy such that packet drops
may be minimized.

By splitting LLC into two partitions, \citeauthor*{sprabery_novel_2017}~\cite{sprabery_novel_2017} attempts to
protect sensitive applications against cache side-channel attacks. Sensitive workloads
are assigned to one, while other, insensitive, workloads share the other. The cores
assigned to the sensitive partition will upon context switching perform a ``state cleansing''.
This means that the LLC used by the previous execution context will be flushed before the
new context begins execution. This approach protects sensitive data at the cost of performance.

\subsubsection{Reactive Approaches}

\revised{Reactive approaches adjust resource partitioning schemes in response
    to changes in workload behaviors and system metrics. They do not attempt to 
    predict future resource requirements, but rather react to changes in the system 
    state as they occur.}

\citeauthor*{navarro-torres_balancer_2023}~\cite{navarro-torres_balancer_2023} identify changes
in program phase by observing cache and memory bandwidth usage behaviors. Such changes trigger
reevaluations of the current allocation scheme for LLC and memory bandwidth, respectively. The two
resource dimensions are controlled by separate heuristic schemes, that work together to
balance resource congestion. This controller is called \emph{Balancer}, which aims to improve 
overall system performance and thread fairness.

\emph{HyPart}~\cite{park_hypart_2018} is a dynamic memory bandwidth 
partitioning controller with a focus on fairness proposed by \citeauthor*{park_hypart_2018} that uses 
workload profiling as a base for its explorative approach to resource partitioning.
Profiling workload behavior allows the scheme to eliminate potential states from
the state space it uses to iteratively and heuristically converge on some optimum
partitioning scheme. The process is repeated should the state of the system change.
By expanding upon this work, the authors later propose
\emph{CoPart}~\cite{park_copart_2019}. In this extended work which we previously 
highlighted, the controller is
able to optimize for performance in addition to fairness. It is also extended
to partition and manage LLC allocation instead of solely managing memory bandwidth.

\citeauthor*{xu_dcat_2018}~\cite{xu_dcat_2018} propose \emph{dCat}, 
a dynamic LLC partitioning controller that aims to improve the 
performance of system workloads. This is done through heuristic workload phase detection 
based on cache access, instruction retirement, and instructions per cycle (IPC) metrics.
Workloads are classified into the classes \emph{reclaim}, \emph{receiver}, \emph{donor}, 
\emph{keeper}, \emph{streaming}, and \emph{unknown} based on their current cache utilization
behavior. Allocation changes are triggered by applications transitioning between states 
due to changes in cache access behavior.

\citeauthor*{pons_cache-poll_2022}~\cite{pons_cache-poll_2022} aim to improve system
fairness and performance by identifying cache-polluting processes and limiting their
ability to impact cache-sensitive applications with \emph{Cache-Poll}. Their approach is
a heuristic algorithm that classifies applications based on system metrics into one of
three classes. Cache sensitive, mildly cache sensitive, and cache insensitive. 
Based on the said classification, Cache-Poll then partitions available LLC in order to improve 
performance and fairness.

By building upon their work presented in \cite{farina_assessing_2022}, 
\citeauthor*{farina_enabling_2023}~\cite{farina_enabling_2023} propose a heuristic
for estimating and guaranteeing memory access for critical cores while not
significantly sacrificing the performance of non-critical ones. Their approach is
based on observation and analysis of the un-core metric read pending queue (RPQ)
found on Intel processors.

\citeauthor*{roy_satori_2021}~\cite{roy_satori_2021} increase system throughput and
fairness of multiple BE jobs competing over system resources. They do this by proposing \emph{Satori}, a 
resource partitioning controller utilizing Bayesian optimization to optimize cache and memory
bandwidth allocations based on a novel objective function using system metrics.

By observing a tenant VM's ability to significantly reduce the performance of co-located 
VMs, \citeauthor*{chen_alita_2020}~\cite{chen_alita_2020} argue that a dynamic 
resource management system is needed to improve fairness among tenants. Their proposal
called \emph{Alita} identifies resource polluters and limits their availability to
resources that are under pressure. The identification of such polluters is done solely
through the use of hardware and OS-level metrics.

\citeauthor*{xiang_dcaps_2018}~\cite{xiang_dcaps_2018} approach the problem of
cache partitioning by searching the large space of possible partitioning combinations for
a near-optimal configuration. They do this by proposing an algorithm based on the
simulated annealing optimization scheme that optimizes workload performance.
This scheme is called \emph{dynamic allocation with partial sharing} (\emph{DCAPS}), and
operates in user space, using only system metrics.

Intel's Data Direct I/O (DDIO) technology enables the transfer of data directly between network
controllers and LLC, completely bypassing main memory and increasing performance. However, this
may introduce additional LLC congestion. In an attempt to mitigate this, 
\citeauthor*{yuan_dont_2021}~\cite{yuan_dont_2021} propose \emph{IAT}. 
IAT is a heuristic cache partitioning scheme that resembles a Mealy 
finite state machine (FSM). Workloads are classified into states that map onto states of the FSM.
Based on system metrics, workloads may change state and thus trigger LLC reallocation. 
IAT is DDIO aware and may allocate part of the LLC exclusively to DDIO to improve
system performance.

Through observation of system metrics, \citeauthor*{yi_mt2_2022}~\cite{yi_mt2_2022} aim
to improve system performance on platforms with non-volatile memory (NVM) by regulating 
memory bandwidth with their \emph{MT$^2$}
controller. The kernel-space controller allows administrator control of its 
throttling groups, which are based on the Linux kernel's cgroups feature. By continuously
updating the bandwidth throttling settings based on application bandwidth usage,
overall system performance may be improved.

To protect against cache timing channel attacks between hardware threads, applications,
or VMs, \citeauthor*{yao_leveraging_2019} proposes a system that monitors cache
behavior and heuristically identifies pairs of workloads with suspiciously correlated
activity~\cite{yao_leveraging_2019}. Any suspected spy-victim pair are separated by 
strictly separating their available cache.

\subsubsection{Combined Approaches}

By leveraging both proactive and reactive strategies, combined approaches are able
to create approximate partitioning schemes that may be fine-tuned by observing
system metrics.

By classifying workloads based on cache usage behavior,
\citeauthor*{li_pfa_2022}~\cite{li_pfa_2022} propose a performance and fairness-aware
LLC partitioning scheme called \emph{PFA}. Classification is performed through the observation 
of cache hit rates of retired CPU instructions, after which a prediction of LLC requirements is made.
During operation, performance slack is monitored and partitioning is adjusted accordingly. 
Improvements are observed in both performance and 
fairness using their controller, but it is noted that further improvements should 
be achievable by tuning memory bandwidth in combination with LLC partitioning.

With a clear focus on fairness, \citeauthor*{garcia-garcia_lfoc_2019}~\cite{garcia-garcia_lfoc_2019}
propose \emph{LFOC}, a controller that classifies workloads by their cache behavior.
The classes defined are \emph{light sharing}, \emph{streaming}, and \emph{sensitive}. 
LLC is then partitioned accordingly using a heuristic cache-clustering algorithm to increase 
overall system fairness. This work is later extended in \emph{LFOC+}~\cite{saez_lfoc_2022} 
where an improved pair-clustering strategy is used to further improve system fairness.

By dividing the LLC into two partitions, \citeauthor*{pons_improving_2018}~\cite{pons_improving_2018}
attempt to improve system performance and fairness. Workloads are profiled offline and classified as either 
cache sensitive or cache insensitive based on how they respond to changes in cache availability, 
after which they are assigned to the corresponding LLC
partition. By dynamically ``moving the goalpost'', i.e. shrinking one partition and increasing 
the other, they aim to decrease system congestion. Their heuristic approach is called
\emph{critical-aware} or simply referred to as \emph{CA}. This work is later extended in
\emph{CPA}~\cite{pons_phase-aware_2020} which uses instructions
per cycle and cache hit rate metrics to determine the phase of a workload and allocates 
LLC accordingly. Here, LLC partitioning is not limited to just two partitions.

Through their use of deep reinforcement learning (DRL), 
\citeauthor*{chen_drlpart_2021}~\cite{chen_drlpart_2021} present a resource allocation
controller called \emph{DRLPart} that partitions the resources of a system among 
workloads using a reward prediction model. The goal of the allocations is to
reduce resource congestion and increase workload throughput. To validate its effectiveness, 
a comparison with two older controllers~\cite{patel_clite_2020, xiang_dcaps_2018} is carried out.

By allowing the compiler to create hints about the cache behavior of programs, 
\citeauthor*{chatterjee_com-cas_2023}~\cite{chatterjee_com-cas_2023} are able to
make proactive, just-in-time, LLC allocation decisions. Their proposed scheme, 
\emph{Com-CAS}, makes use of a compilation component based on LLVM and a
runtime component that works in unison to achieve improved system performance.

With the goal of mitigating resource under-utilization in large-scale Kubernetes clusters,
\citeauthor*{zhang_zeus_2021} detail their resource management scheme \emph{Zeus}~\cite{zhang_zeus_2021}.
Zeus intercepts Kubernetes API requests and modifies allocation requests in accordance with its 
observed node-level metrics. On each node, LLC is split into two partitions. One for LC workloads,
and one for BE jobs. This allows the controller to minimize interference and thus
increase average, cluster-wide, CPU utilization from $15\%$ to $60\%$.

To improve the performance of accelerated ML workloads, \citeauthor*{zhu_kelp_2019} leverage
non-uniform memory access (NUMA) node subdomains in their \emph{Kelp}~\cite{zhu_kelp_2019} runtime. 
Each system socket is split into two subdomains, one for high-priority ML workloads and one for low-priority tasks.
This separation in combination with heuristic resource partitioning reduces task interference such that ML workload
throughput is increased.

\subsubsection{Workload Clustering Approaches}

Approaches that cluster workloads attempt to group workloads by behavior so that
allocation of resources may be performed in a generalized fashion. 

\emph{Themis}~\cite{tang_themis_2023} by \citeauthor*{tang_themis_2023} increases QoS and fairness
of guest VMs by observing their memory bandwidth usage behavior. The VMs are classified into
groups using the Euclidean K-Means algorithm, where the number of groups equal the number of 
available COS on the system. The controller then proactively partitions LLC and MB by following
a heuristic algorithm. The classification and partitioning scheme is reevaluated over time by
continuously observing VM bandwidth usage. The controller may also react to saturation of 
the memory bus by moving VMs to other NUMA nodes if available.

By observing a core's LLC cache miss percentage and number of retired instructions, 
\citeauthor*{gifford_dna_2021}~\cite{gifford_dna_2021} classify workloads into clusters
depending on the program phase they are deemed to currently be in.
These clusters are then allocated LLC ways and memory bandwidth\footnote{
    Memory bandwidth is allocated using \emph{MemGuard}~\cite{yun_memguard_2013}, a software-based approach.
} in order to maximize the performance of a system based on a heuristic algorithm. 
They call their approach \emph{DNA}. A deadline-aware version that is able to take a real-time workload's 
execution deadlines into consideration called \emph{DADNA} is also presented. Their experimental 
evaluation of the controller only compares DNA and DADNA against \cite{xu_holistic_2019}, 
another controller previously proposed  by the same authors.

With the goal of increasing overall system performance, 
\citeauthor*{xiang_emba_2019}~\cite{xiang_emba_2019} propose \emph{EMBA}. EMBA is a controller 
that clusters workloads using a heuristic hierarchical clustering algorithm based on 
their usage of the memory bus. This classification is then
used to heuristically and iteratively achieve a memory bandwidth throttle scheme 
that improves overall performance.

With system fairness as their goal, 
\citeauthor*{selfa_application_2017}~\cite{selfa_application_2017} propose a set of 
LLC allocation policies that are based on workload clustering using the K-Means 
algorithm~\cite{hartigan_algorithm_1979}. The classification is done based on the number of 
stalled CPU cycles due to L2 cache misses. Then, the LLC is partitioned accordingly 
which results in an improvement of overall system fairness.

By clustering workloads using a support vector machine (SVM) based machine learning model, 
\citeauthor*{qiu_machine-learning-based_2022} are able to classify workload behavior using
only system metrics. This is the first phase of their \emph{Clustering-and-Allocation}
(C\&A)~\cite{qiu_machine-learning-based_2022} LLC partitioning scheme. Using the
aforementioned workload classification as input, a Bayesian optimization-based approach
is used to quickly and efficiently reach a near-optimum partitioning scheme.
In their own evaluation, C\&A is found to increase throughput by up to $26.2 \%$
in a best-case scenario.

\subsubsection{Feature Abstractions}

Due to the nature of hardware features such as Intel CAT, they are limited to bare-metal
deployments with direct access to hardware. In an attempt to bring LLC allocation to virtualized
environments, \citeauthor*{xu_vcat_2017}~\cite{xu_vcat_2017} propose \emph{vCAT}. vCAT is a system that
exposes virtual cache partitioning to VMs, which is later mapped to physical LLC partitions 
on the host. By the extension of the functionality of hypervisors, guests are able to give hints
regarding their workloads' requirements, while the host remains in control.

\subsubsection{Summary}

Hardware QoS enforcement features may be used to great effect and tuned in many ways even when
only system metrics are observed. Throughput may be increased by $26.2 \%$~\cite{qiu_machine-learning-based_2022} or
$18.5\%$~\cite{chen_drlpart_2021}, performance increased by $7.1 \%$~\cite{navarro-torres_balancer_2023},
or fairness improved by $44 \%$~\cite{pons_cache-poll_2022} compared to an unmanaged system.

Many publications classify workloads with
regard to their utilization and sensitivity to a shortage of shared system 
resources. This categorization is then often used to dictate how resources are 
partitioned among workloads. Other papers try to predict how changes in
resource allocation would impact the performance of workloads through 
machine learning and mathematical approaches. Conversely,
reactive controllers have been proposed that continuously profile the 
workloads and reactively adjust resource allocations as requirements
change over time.
Additionally, publications focused on security~\cite{liu_catalyst_2016}, 
compiler inserted cache behavior hints~\cite{chatterjee_com-cas_2023},
auctioning-based partitioning~\cite{funaro_ginseng_2016}, and
LLC management on the data-center scale~\cite{shahrad_provisioning_2021},
and more have been discussed.

\subsection{Application Aware Approaches} \label{sec:application-aware}

\begin{figure}[htbp]
    \centering
    \small
    \includegraphics{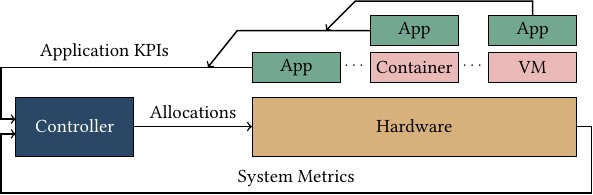}
    \caption{\revised{Illustration of application-aware approaches and how they rely on application KPIs, sometimes in addition to system metrics.
    KPIs may include request throughput, average latencies, tail latencies, etc.}}
    \label{fig:application-aware}
    \Description{Illustration of application-aware approaches and how they rely on application KPIs, sometimes in addition to system metrics.}
\end{figure}

To better understand how shared resource allocation affects the real-world performance
of workloads, controllers may monitor key performance indicators (KPIs) of applications, 
\revised{see \figurename~\ref{fig:application-aware}}. Such metrics
may include request throughput, average latencies, tail latencies, and more. Every workload's KPIs are different
and manual configuration of metric probes is therefore often required. Resource management 
controllers need to be aware of how to measure an application's metrics, as well as their target values and
the importance of one metric compared to another. Many workloads have service-level objectives (SLOs) 
to adhere to which allow application-aware resource managers to minimize the resource 
availability of such workloads while ensuring they still operate within their targets. 
Thus, freeing more resources to maximize the performance of other co-located workloads.

We group application-aware approaches into four categories: Proactive, reactive, combined, 
and performance slack management approaches. See \tablename~\ref{tbl:application-aware}.
    
\begin{table*}[t]
    \centering
    \small
    \caption{Publications leveraging both system and application metrics.}
    \label{tbl:application-aware}
    \begin{tabular}{ccclll}
        \toprule
        \textbf{Ref.} & \textbf{CA*} & \textbf{MBL*} & \textbf{DC*} 
            & \textbf{Optimization} & \textbf{Proposed Technique} \\
        
        \midrule \multicolumn{6}{c}{Proactive Approaches} \\ \midrule
        
            \cite{jeatsa_casy_2022} & \checkmark &  & Cloud
                & Performance & ML Resource Management Controller for FaaS \\ 
            \cite{stewart_performance_2022} & \checkmark &  & Generic
                & Performance & ML Short-term LLC Allocation Controller \\ 
            \cite{noll_accelerating_2018} & \checkmark & & DBMS
                & Performance & LLC Management for DBMS \\ 
                
        \midrule \multicolumn{6}{c}{Reactive Approaches} \\ \midrule
        
            \cite{nikas_dicer_2019} & \checkmark & & Generic
                & Utilization & Heuristic LLC Partitioning Scheme \\
            \cite{chintapalli_restrain_2022} & \checkmark & \checkmark & VNFs
                & Performance & Heuristic VNF Resource Management \\
            \cite{duan_improving_2021} & \checkmark & & Generic
                & Performance & ML Classification and Heuristic Scheduling \\
            \cite{duan_improving_2022} & \checkmark & & Generic
                & Performance & ML Classification and Heuristic Scheduling \\
            \cite{patel_clite_2020} & \checkmark & \checkmark & Cloud
                & Performance & Bayesian Optimization Based Algorithm \\ 
            \cite{pons_effect_2022} & \checkmark & \checkmark & Generic
                & Performance & Analysis of App Sensitivity to Shared Resources \\ 
            \cite{gureya_generalizing_2021} & & \checkmark & Cloud
                & Performance & Heuristic Multi-Socket Memory Management \\ 
            \cite{papadakis_improving_2017} & \checkmark & & Generic
                & Performance & Heuristic LLC Partitioning Scheme \\
            \cite{qin_locpart_2017} & \checkmark & & Generic
                & Performance & Heuristic LLC Partitioning Scheme \\
                
        \midrule \multicolumn{6}{c}{Combined Approaches} \\ \midrule
        
            \cite{qiu_slo_2021} & \checkmark & \checkmark & Generic
                & Performance & Profiling \& Feedback Based Management \\
            \cite{nejat_coordinated_2020} & \checkmark & & Generic
                & Energy Efficiency & Modeling of Workload Behavior \\ 
            \cite{zhao_rhythm_2020} & \checkmark & & Cloud
                & Utilization & Sub-service LLC Management Scheme \\
            \cite{tootoonchian_resq_2018} & \checkmark & & VNFs
                & Utilization & Greedy Profiling Based LLC Allocation\\
            \cite{metsch_intent-driven_2023} & \checkmark & & Cloud
                & Performance & ML and Intent-driven Resource Management \\
            \cite{li_rldrm_2020} & \checkmark & & VNFs
                & Utilization & Deep Reinforcement Learning LLC Management \\
            \cite{zhang_libra_2021} & \checkmark &  & Generic
                & Performance & Heuristic Memory Bandwidth Management \\ 
            \cite{chintapalli_ravin_2023} & \checkmark & \checkmark & VNFs
                & Utilization & Heuristic VNF and Shared Resource Allocation \\  
            \cite{xu_holistic_2019} & \checkmark & & RT
                & Performance & Heuristic LLC and BW Management Algorithm \\
                
        \midrule \multicolumn{6}{c}{Performance Slack Management Approaches} \\ \midrule
        
            \cite{chen_parties_2019} & \checkmark & & Cloud
                & Utilization & Heuristic Slack Management Algorithm  \\
            \cite{zhu_dirigent_2016} & \checkmark & & Generic
                & Utilization & Execution Time Prediction Algorithm \\
            \cite{lo_heracles_2015} & \checkmark & & Cloud
                & Performance & Heuristic Slack Management Algorithm  \\ 
            \cite{nejat_cooperative_2022} & \checkmark & & Generic
                & Energy Efficiency & Heuristic Slack Management Algorithm  \\
            \cite{pang_adaptive_2021} & \checkmark & & Cloud
                & Energy Efficiency & Multi-level Resource Management Scheme \\
            \cite{chen_avalon_2019} & \checkmark & & Generic
                & Utilization & Runtime Prediction and LLC Management \\
                
        \bottomrule
          
        \multicolumn{6}{c}{\scriptsize *Abbreviations --- CA: Cache Allocation, MBL: Memory Bandwidth Limiting, DC: Discussion Context.} 
            
    \end{tabular}
\end{table*}

\begin{figure}[htbp]
    \centering
    \includegraphics{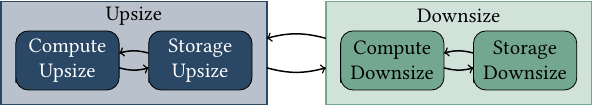}
    \caption{\revised{Illustration of the iterative resource adjustment scheme of PARTIES by 
    \citeauthor{chen_parties_2019}~\cite{chen_parties_2019}.}}
    \label{fig:parties}
    \Description{Illustration of the iterative resource adjustment scheme of PARTIES by 
    \citeauthor{chen_parties_2019}.}
\end{figure}

\revised{As one of the more recognized and important works in the category, 
\emph{PARTIES}~\cite{chen_parties_2019} is a QoS-aware resource management 
controller that allocates and tunes LLC, CPU cores, CPU frequency, and disk I/O
to improve overall performance.  
It works with a single LC workload co-located with multiple BE jobs that initially 
are allocated equal amounts of resources. The controller then observes each application's
tail latency to determine its performance slack in relation to a predefined QoS target. The applications with small or
negative slack is selected for \emph{upsizing}, where the controller iteratively
tries to increase allocations one resource at a time until the slack is increased.
Should an increase in an allocation not lead to increased slack, the change is reverted.
This is similarly repeated for applications with large slack, but in the opposite
direction through \emph{downsizing}. This process is repeated while applications
fulfill predefined criteria for either upsizing or downsizing and is illustrated in \figurename~\ref{fig:parties}.
In the event that the controller is unable to find a configuration that satisfies all
applications' SLOs, it will select an appropriate application for node migration~\cite{chen_parties_2019}.}

\subsubsection{Proactive Approaches}

By predicting the resource requirements of workloads~\cite{jeatsa_casy_2022,stewart_performance_2022}, 
or by observing workload characteristics to identify beneficial static 
configurations~\cite{noll_accelerating_2018}, proactive approaches 
improve system goals ahead of time.

\emph{CASY}~\cite{jeatsa_casy_2022} by \citeauthor{jeatsa_casy_2022} 
is a cache allocation controller for function-as-a-service (FaaS) systems that uses machine learning to build 
a cache usage profile of functions with regard to their input values. This allows
the controller to tune functions' allocation of cache by examining the incoming input 
and thus improve system-wide performance.

With the use of deep learning, \citeauthor*{stewart_performance_2022}~\cite{stewart_performance_2022}
profile web services' cache usage characteristics. This allows them to propose their short-term
cache allocation controller that observes response times of web services and temporarily allocates
additional LLC to particularly slow requests in order to minimize SLO infringement.

Database management systems (DBMS) are required to handle large amounts of queries and data, but all
operations are not equal in cache sensitivity and efficiency as noted by 
\citeauthor{noll_accelerating_2018}~\cite{noll_accelerating_2018}.
They find that certain memory-intensive operations such as a column scan are insensitive to available cache, while
simultaneously thrashing existing cache contents. 
Thus, they propose allocating the minimum possible cache to worker threads that handle such queries which 
reduces overall cache pollution and thus increases system performance by, in their evaluation, up to $38 \%$.

\subsubsection{Reactive Approaches}

Application-aware reactive approaches monitor application metrics and identify required adjustments
to resource partitioning schemes as application load and type of load change over time. They react
by applying corrective allocation adjustments that allow optimization of the controllers' respective goals.

Through observation of both application metrics and system metrics, \emph{DICER}~\cite{nikas_dicer_2019} 
classifies workloads based on their sensitivity to LLC availability. Additionally, it monitors MB 
saturation and detects workload phase changes by identifying changes in LLC utilization. 
The controller proposed by \citeauthor*{nikas_dicer_2019} uses the gathered data to 
heuristically increase system utilization by improving BE job performance without 
allowing LC workloads to break QoS targets. 

Virtual network functions (VNFs) often have strict performance guarantees to meet.
\citeauthor*{chintapalli_restrain_2022}~\cite{chintapalli_restrain_2022} argue that to be
able to co-locate multiple VNFs with other services while ensuring QoS, isolation of 
resources is crucial. They present \emph{RESTRAIN},
a dynamic LLC and MB allocation scheme that continuously
classifies VNFs into one of four categories using a modified technique borrowed from 
\citeauthor*{park_copart_2019}~\cite{park_copart_2019}. The categories used are ``donor'', ``receiver'', 
``donor and receiver'', and ``none''. A donor is a workload that is able to release 
a certain resource and still meet QoS, while receivers need more resources to meet QoS.
Donors and receivers are workloads that require more of one resource while being in
excess of another. Workloads that are at a  resource allocation are 
classified as none. The controller builds upon and is compared 
to ResQ~\cite{tootoonchian_resq_2018}.

With large scale data-centers in mind, \citeauthor*{duan_improving_2021}~\cite{duan_improving_2021} propose a
machine learning-based workload classification model. This model is later used by a local search 
server re-consolidation algorithm. The controller utilizes LLC partitioning and core allocation 
to guarantee workload performance during re-consolidation from overloaded machines. This work is 
continued and extended by the same authors in \citeauthor*{duan_improving_2022}~\cite{duan_improving_2022}.

With the goal of increasing system utilization by co-locating a greater number of BE jobs without breaking
LC service SLOs, \citeauthor*{patel_clite_2020}~\cite{patel_clite_2020} propose \emph{CLITE}. CLITE is a
memory bandwidth and LLC partitioning controller that uses Bayesian optimization to find an optimal
configuration of the system. In their case, the controller maximizes the performance of BE jobs while
LC applications still meet their QoS requirements.

Identifying the most critical resource of a workload at runtime can be difficult.
\citeauthor*{pons_effect_2022}~\cite{pons_effect_2022} try to show how different types
of workloads behave when constrained by different kinds of resources. This gives an insight
into how resource allocation features behave when used in conjunction. For example, they show 
that a reduction in LLC availability may increase the memory bandwidth used by a process.
This allows them to present a number of key takeaways and guidelines that may 
help guide cloud operators to make more efficient use of their hardware by reducing 
resource allocation to insensitive workloads.

\citeauthor*{gureya_generalizing_2021}~\cite{gureya_generalizing_2021} recognize that Intel's MBA
is limited to single socket control only due to the details of its implementation. In an attempt
to bring holistic memory bandwidth allocation to multi-socket systems, they propose \emph{BALM}.
BALM is a controller that combines a cross-socket page migration scheme with knowledge of
workloads' SLO targets. This allows the controller to achieve better overall system performance
through memory bandwidth partitioning and socket migration of workloads 
while maintaining SLO targets.

By allocating an exclusive cache partition to a high-priority workload while other applications
share the rest, the QoS of the high-priority workload can be maintained. This is what
\citeauthor*{papadakis_improving_2017} propose with their heuristic \emph{Dynamic Cache-Partitioning QoS}
scheme~\cite{papadakis_improving_2017}. The controller observes the performance of the high-priority
workload and adjusts its LLC allocation such that QoS is maintained while maximizing LLC available
to remaining system workloads.

\citeauthor{qin_locpart_2017} approach LLC partitioning by allocating each workload a
separate exclusive piece of LLC in \emph{Locpart}~\cite{qin_locpart_2017}. By monitoring 
the QoS of LC workloads, LLC may be redistributed
to or from LC workloads to BE jobs in such a way that LC services meet their performance targets
and BE job performance is maximized. A limitation of the approach is that the number of available
COSs dictates the maximum number of total workloads assigned to a system.

\subsubsection{Combined Approaches}

Combined application-aware approaches utilize a combination of proactive and reactive techniques
to achieve their goals. Intuitively, this means they perform some ahead-of-time prediction of resource
requirements, perhaps through offline application profiling. Then, at runtime, the workloads are 
monitored and minor corrective adjustments are performed
as load and workload behavior changes over time.

With the goal of increasing system performance while maintaining LC workload SLO requirements,
\citeauthor*{qiu_slo_2021}~\cite{qiu_slo_2021} propose \emph{CoCo}. CoCo is an SLO-aware 
LLC and memory bandwidth partitioning controller that profiles workloads' sensitivity to
such scarce resources. With this knowledge, it dynamically allocates resources based on available SLO 
overhead to increase overall system utilization and performance.

\citeauthor*{nejat_coordinated_2020} have total system efficiency in mind when
proposing their resource management controller~\cite{nejat_coordinated_2020}. By observing the SLOs
and metrics of workloads, their controller is able to reduce power consumption while still meeting those 
QoS requirements. This is achieved by modeling the cache and power usage behavior of workloads 
and then partitioning LLC and scaling core voltages to reduce energy consumption while still allowing 
applications to meet their intended targets.

By gathering multiple co-located parts of the same LC service into a new abstraction called a Servpod, 
\citeauthor*{zhao_rhythm_2020}~\cite{zhao_rhythm_2020} are able to improve system utilization while 
ensuring that LC workloads meet their SLOs. This is
achieved by offline profiling and classification of Servpods based on their contributions to overall system 
tail-latency and sensitivity to shortage of shared scarce resources. They call their scheme \emph{Rhythm}, 
which partitions LLC, 
manages core voltages, etc., with the goal of increasing utilization without LC SLO breakage.

By using the result from offline profiling of VNFs prior to deployment, 
\emph{ResQ}~\cite{tootoonchian_resq_2018} is able to make greedy decisions regarding 
LLC allocation optimizations. The controller, proposed by
\citeauthor*{tootoonchian_resq_2018}, ensures VNF QoS requirements are met while maximizing 
the performance of co-located BE workloads. 

To allow workload owners to specify SLO targets instead of resource allocation requests in the cloud,
\citeauthor*{metsch_intent-driven_2023}~\cite{metsch_intent-driven_2023} propose their 
\emph{intent driven orchestrator}. The system learns from application behavior and how changes
in the allocation of resources such as LLC affect SLO metrics. This allows the controller to predict
how changes in resource allocation will affect application performance and make resource partitioning
decisions accordingly.

Due to strict SLO requirements, VNFs are often placed on underutilized 
machines so that their performance can be guaranteed. 
\citeauthor*{li_rldrm_2020}~\cite{li_rldrm_2020} propose \emph{RLDRM} in an attempt to improve 
utilization of such machines. RLDRM is a closed-loop deep reinforcement learning-based 
LLC partitioning controller. By learning from the behavior of workloads the controller can make sure the 
high-priority VNFs have enough LLC ways to meet their SLO requirements while allowing a greater number of 
BE jobs to consume the remaining resources. 

After identifying the shortcomings of Intel's MBA, \citeauthor*{zhang_libra_2021}~\cite{zhang_libra_2021} 
propose a custom memory bandwidth partitioning mechanism they call \emph{LIBRA}. By custom 
manipulation of the read pending queue, finer-grained control is achieved than what is available through
MBA. LIBRA still leverages CAT as a mechanism of LLC partitioning.

Through prior profiling of VNF sensitivity to cache and memory bandwidth
constraints, \citeauthor*{chintapalli_ravin_2023}~\cite{chintapalli_ravin_2023} propose \emph{RAVIN}.
The heuristic VNF orchestrator schedules VNFs across machines while considering available cache and 
memory bandwidth such that throughput guarantees may be met. This is done using the previously gathered 
profiling information which provides the orchestrator with minimum resource requirements depending on
desired throughput guarantees. This allows RAVIN to efficiently utilize available hardware through
more efficient packing of VNFs.

To ensure that the worst-case execution time (WCET) of real-time workloads 
stays within requirements, \citeauthor*{xu_holistic_2019} propose \emph{CaM}~\cite{xu_holistic_2019}. The three-phase heuristic
algorithm groups workloads based on their sensitivity to resource shortage before it load-balances
them across the cores of a system. LLC and memory BW (using MemGuard~\cite{yun_memguard_2013}) are then allocated 
such that WCET targets are fulfilled.

\subsubsection{Performance Slack Management Approaches}

Management of performance slack is a concept commonly seen in the context of real-time systems.
When operating under completion deadlines, workloads may be tuned such that they may operate 
``ahead of schedule'' which allows performance to be prioritized elsewhere at times of need. This 
notion of slack is used by several application-aware approaches utilizing hardware QoS enforcement.

With knowledge of an LC workload's QoS requirements and prior offline profiling, \emph{Dirigent}~\cite{zhu_dirigent_2016} 
may manage LLC in such a way that co-located BE jobs' performance can be maximized and thus increase overall
system utilization. \citeauthor*{zhu_dirigent_2016} approach the problem by predicting the remaining runtime of
workloads using data gathered during the prior profiling. By comparing how it correlates to set deadlines, resources
may be adjusted dynamically so that the performance of LC workloads meets set targets while 
allowing an increased amount of resources to co-located jobs.

In one of the first works utilizing platform features, \citeauthor*{lo_heracles_2015} propose 
\emph{Heracles}~\cite{lo_heracles_2015}, a resource partitioning controller that monitors QoS metrics of an LC workload. By observing that
LC workload's performance slack, its resource allocations may be adjusted such that just enough resources are
available to meet its SLO targets. This allows the remaining resources to be distributed among BE jobs, 
thus improving system utilization and the performance of those services.

By being aware of workload QoS targets, a controller may locate LLC resources
in a way that decreases the overall power consumption of a system through the management
of performance slack. This is what \citeauthor*{nejat_cooperative_2022}~\cite{nejat_cooperative_2022}
propose with their \emph{Cooperative Slack Management} scheme. It identifies opportunities to
generate slack with minimal use of energy. This slack may later be consumed at times of
less power-efficient operation and thus increase energy efficiency over time. 

\emph{Sturgeon}~\cite{pang_adaptive_2021} is a multi-level resource management system that manages BE
and LC workload placement at the cluster level in addition to providing machine-level 
resource management with a focus on achieving higher performance while not increasing energy consumption.
They propose using machine learning models to predict both the performance and power consumption of
heuristic changes to resource partitioning based on performance slack. This is to ensure services
stay within SLO requirements and that the system does not breach power usage constraints.

\emph{Avalon}~\cite{chen_avalon_2019} is a resource management scheme proposed by \citeauthor*{chen_avalon_2019} 
that attempts to improve system utilization by providing ``just enough'' resources to LC services while
ensuring QoS targets are met. This is done by predicting the resource requirements for each incoming LC workload
query using an ML model and allocating resources appropriately. The execution time slack of queries is monitored
and heuristic adjustments to resource allocation are made if necessary to meet QoS targets. This improves utilization
by providing co-located BE jobs with increased resources while ensuring LC QoS.

\subsubsection{Summary}

Being application-aware means a higher configuration overhead for controllers managing a 
heterogeneous set of workloads. However, it does allow controllers to optimize the resource 
partitioning of a system while maintaining workload SLO targets. Publications covered
in this section have seen improvements in LC workload throughput by $31.7 \%$~\cite{zhao_rhythm_2020},
system utilization by $28.9 \%$~\cite{chen_avalon_2019}, and serverless function 
execution time by $11 \%$~\cite{jeatsa_casy_2022}. It is therefore clear that application-aware 
management of QoS enforcement features may greatly benefit execution environments not already
configured to limit congestion over shared resources.

The problem of shared resource management is approached in the context of virtual 
network functions by \cite{li_rldrm_2020, chintapalli_restrain_2022, tootoonchian_resq_2018, chintapalli_ravin_2023}.
It is common that such functions are constrained by strict QoS requirements, and are thus 
traditionally deployed on machines with relatively low total utilization~\cite{li_rldrm_2020}.
By guaranteeing scarce resource access, utilization of such machines may be greatly improved
for reduced cost of ownership and increased efficiency.

By observing LC workload deadlines and by efficiently managing performance slack, 
efficiency~\cite{nejat_cooperative_2022, pang_adaptive_2021}, 
server utilization~\cite{zhu_dirigent_2016, chen_parties_2019, chen_avalon_2019},
or system performance~\cite{lo_heracles_2015} may be improved. This is done by identifying
times at which slack may be increased at a low cost such that it may be consumed at a later
time when the cost is higher.

\section{Discussion} \label{sec:discussion}

Researchers have over the years leveraged the quality of service
enforcement features of modern processors to propose a wide range of
state-of-the-art proposals that optimize system performance, utilization,
fairness, power consumption, and much more. As illustrated in
\figurename~\ref{fig:hardware-year}, the interest and utilization of such features 
remains high as more and more creative use cases are found. 
This is not exclusive to academic circles but includes the industry actors, 
as indicated by the many collaborative works published between the academic and private sectors.
However, the taxonomy presented in this survey shows a distinct lack of publications
presenting modern, generally applicable, general-purpose cloud computing approaches.
This is especially true when considering papers
originating from the public clouds and their providers, who stand to gain major
improvements in operational efficiency through such research. 

We have found a few exceptions to this trend. Firstly,
CacheSlicer~\cite{shahrad_provisioning_2021} is a proof-of-concept
of an LLC-aware extension of the Microsoft Azure virtual machine scheduler.
Secondly, Zeus~\cite{zhang_zeus_2021} attempts to improve upon Kubernetes resource
management and increase cluster utilization at NetEase. Finally, the intent-driven
orchestrator~\cite{metsch_intent-driven_2023} tries to learn how changes in resource
allocations affect service metrics and then make appropriate adjustments.

This shortage of published material may suggest one of a number of possible scenarios: 
\begin{enumerate}
    \item \label{scen1} Larger corporations and cloud providers do not see potential in the available
        hardware-based QoS platform features and do thus not wish to invest resources
        into exploring a dead end.
    \item \label{scen2} The private sector is researching, investing in, and developing
        efficient ways of utilizing recent QoS features behind closed doors. Thus gaining
        a competitive advantage by not sharing their findings.
\end{enumerate}
Given the large number of published works surveyed in this paper that show significant 
improvements to be gained through the effective usage of QoS features, we find scenario 
(\ref{scen1}) highly unlikely. When considering that chip manufacturers have
continued to develop and improve, generation by generation, upon a feature that
potentially none of their major corporate customers utilize, scenario (\ref{scen2})
seems increasingly more likely to reflect reality.

\subsection{Privacy Concerns of Private and Public Clouds}

The type of cloud operated has a big impact on the type of approaches usable in a 
given scenario. A private cloud, hosted on-premise, running only software written in-house
has vastly different requirements than a cloud running customer software. In the first scenario,
cloud operators are able to monitor applications in any way imaginable. This includes prior-to-deployment 
offline profiling of software. This allows resource management controllers with full application-aware
integration to be employed, as leveraged by the intent-driven 
orchestrator~\cite{metsch_intent-driven_2023}. However, companies hosting other organizations' software
through infrastructure-as-a-service (IaaS) solutions are required to take additional security 
measures into account as detailed by 
\citeauthor*{shahrad_provisioning_2021}~\cite{shahrad_provisioning_2021}. In such scenarios, as
in the context of public cloud providers, there is no alternative but to treat customer
software as black boxes. Nonetheless, managed platform-as-a-service (PaaS) offerings could
provide software add-ons allowing customers to provide a resource manager with 
application-level metrics and targets. Such solutions would make customer software
a little more ``grey'' without becoming an intellectual property protection concern.

Additionally, there may be sensitive data managed by the application that must be handled with discretion
due to legal compliance reasons, such as with the EU's General Data Protection Regulation 
(GDPR)~\cite{noauthor_general_nodate}.

\subsection{Consolidation of Workloads}

Many of the workload consolidation approaches discuss consolidation from the point of view
that workloads are either latency-critical (LC) continuously running services or best-effort (BE) 
jobs. The general idea is to ensure SLOs are met by one~\cite{lo_heracles_2015, zhao_rhythm_2020} 
or more~\cite{chen_parties_2019, patel_clite_2020,nikas_dicer_2019,qin_locpart_2017,
qiu_slo_2021,li_rldrm_2020,zhu_dirigent_2016,chen_avalon_2019} LC services while utilizing additional 
machine resources to pack and maximize the performance of BE jobs. This way, the utilization of
hardware can be increased while performance expectations of services are met. 
While the way different types of workloads are used to improve system utilization is sound, 
the approach assumes a decently static amount of LC traffic and a steady stream of BE jobs 
awaiting processing. They do not consider scenarios where LC traffic may vary enough that replication
scaling is preferred or required, as is common in cloud-based environments. Nor is there any consideration
for the underprovisioning that would emerge if no BE jobs are scheduled. 

Even though the discussion contexts of these papers are that of cloud computing and macro-scale 
data centers, the proposed controllers are inherently micro-scale and operate on the scale of 
a single machine. Therefore, we see a potential for further research, where workload consolidation is
tackled in a more holistic fashion at the macro-scale. Where service replication scaling is considered
and the notion of LC and BE workloads is approached in a less static way.

\subsection{Classification of Workloads}

As may be observed by tables \ref{tbl:black-box} and \ref{tbl:application-aware}, many publications 
include some form of workload classification as part of their proposed
solutions. This is a reasonable way of tackling the resource partitioning problem due to the constraint 
of finite classes of service (CLOS) available in hardware and subsequently the \texttt{resctrl} resource 
management interface. Therefore, workloads are required to be generically grouped together according 
to some scheme. This may range from a simple binary distinction between resource-sensitive and insensitive (e.g. 
\emph{CA}~\cite{pons_improving_2018}) to more sophisticated program phase classification schemes (e.g. 
\emph{dCat}~\cite{xu_dcat_2018}). 

\subsection{QoS Enforcement Feature Strengths and Weaknesses}

Given the numerous studies that have successfully applied both cache allocation and
memory bandwidth limitation to achieve their respective goals (be it performance, fairness,
energy efficiency, utilization, or a combination of the aforementioned goals), it is clear that
the hardware QoS enforcement features surveyed in this article are appropriate tools to utilize
when optimizing for such goals. Additionally, the aforementioned features have also been shown
to be effective in mitigating other issues such as security threats~\cite{liu_catalyst_2016}.

However, the fundamental indirect nature of QoS enforcement features makes them difficult
for practitioners to utilize effectively. In many cases, performance is left on the table
that could be unlocked by integrating QoS enforcement feature management into systems 
such as VM hypervisors and container orchestrators. Such integrations would allow
users who are unaware that their hardware supports such tuning to take full advantage
of their hardware, automatically and effectively.

\section{Research Gap Analysis} \label{sec:gap-analysis}

As we discussed in Section~\ref{sec:discussion}, it is highly likely that public cloud providers make extensive
use of the QoS enforcement features found in recent server processors. However,
our survey of the academic landscape has made it clear that there is space for 
holistic solutions that are able to intelligently manage cache and memory bandwidth 
in the context of multiple machines or even entire data centers. Only a few 
publications approach QoS management in the context of data centers~\cite{
shahrad_provisioning_2021, jeatsa_casy_2022}, which focus solely on
cache partitioning. This leaves room for additional research that 
could improve the utilization of hardware in smaller, on-premises or
edge deployments that generally cannot benefit from the advancements
made behind public cloud providers' closed doors.
A fitting approach would be to extend service orchestrators such as
Kubernetes~\cite{the_kubernetes_authors_kubernetes_2022} with intelligent
cache and memory bandwidth awareness. 

Observing the evaluations listed in \tablename~\ref{tbl:feature-evaluations},
it is notable that no evaluation focuses on the specifics of AMD's QoSE 
implementation of hardware QoS enforcement features. This is significant
as studies have shown inconsistent behavior between
hardware generations~\cite{sohal_closer_2022}, even from the same manufacturer (Intel). Thus, a systematic
evaluation where the behavior of AMD's implementation should be conducted. 
This would be nicely followed by a comparison of the behaviors of the two chip
manufacturers' implementations.

Additionally, it may be observed that there exists no generic solution to controlling systems
spanning different processor models, generations, or even manufacturers. Given the greatly
varying degrees of control and feature support between chips, the adoption of such features
in general resource management systems may be cumbersome due to the wide range of models
to support. Therefore, a general abstraction model may be developed that makes efficient
use of the underlying features while providing an interface that is generic and 
platform-independent. Such an abstraction would lower the friction of feature integration 
into container orchestrators, VM schedulers, and similar platform-heterogeneous systems.

Furthermore, by assuming that most LC cloud workloads are allocated more resources 
than required to account for sudden spikes in load without breaking SLOs, an opportunity 
to improve system-wide efficiency by utilizing this gap between allocation and 
actual utilization may be observed. \revised{A traditional approach to this problem is to
use overcommitment~\cite{cohen_overcommitment_2019}, but this can lead to unpredictable
performance degradation due to increased resource congestion.
A possible hardware-level solution could be to limit workloads that operate with some margin
to their SLOs with intelligent management of QoS enforcement, which should allow net efficiency gains
to be made without impacting the end-user experience.} This concept is in part addressed by 
\citeauthor*{pang_adaptive_2021}~\cite{pang_adaptive_2021} \revised{which indicates the 
feasibility of the technique.} 
However, their approach only rudimentarily utilizes cache partitioning without 
touching memory bandwidth limiting.

Finally, tuning multiple enforcement features simultaneously is difficult
without experiencing undesired side effects. The indirect nature of QoS enforcement
means that the tuning of one workload often affects the performance of co-located workloads.
Thus, there exists an opportunity to develop a solution that allows for the tuning of
multiple features in unison without negatively affecting neighboring workloads. Such a solution
should make use of application awareness to monitor key performance metrics such that
the desired outcome is achieved.

\section{Concluding Remarks} \label{sec:conclusion}

We have conducted a broad survey of the research landscape and identified how academics
are utilizing hardware-based quality-of-service enforcement features implemented in 
commodity x86 processors. Through our use of a reference snowballing approach, a search space 
consisting of $\searchsize$ references was reviewed. Out of which, $\surveysize$ 
papers were identified to meet our specified inclusion criteria. We proposed a novel 
taxonomy of the included papers that structures and classifies the current approaches 
used by scholars. By observation of this taxonomy, conclusions could be drawn regarding the
current general directions of research. 
Furthermore, gaps in current state-of-the-art research were identified and detailed.

\begin{acks}
We would like to thank our anonymous reviewers as well as Ravi Iyer for their input and invaluable feedback. 
Financial support has been provided in part by the Knut and Alice Wallenberg 
Foundation under grant KAW 2019.0352 and by the eSSENCE Programme under 
the Swedish Government’s Strategic Research Initiative.

\end{acks}

\printbibliography

\end{document}